\documentclass[lettersize,journal]{IEEEtran}

\usepackage{graphicx}%
\usepackage{multirow}%
\usepackage{amsmath,amssymb,amsfonts}%
\usepackage{amsthm}%
\usepackage{mathrsfs}%
\usepackage{xcolor}%
\usepackage{textcomp}%
\usepackage{manyfoot}%
\usepackage{booktabs}%
\usepackage{algorithm}%
\usepackage{algorithmicx}%
\usepackage{algpseudocode}%
\usepackage{listings}%

\usepackage{dcolumn}% Align table columns on decimal point
\usepackage{bm}
\usepackage{subfigure}
\usepackage{multirow}
\usepackage{makecell}
\usepackage{booktabs}
\usepackage[numbers,sort&compress]{natbib}

\newtheorem{lemma}{Lemma}

\begin{document}

\title{Distributed Information-theoretical Secure Protocols for Quantum Key Distribution Networks against Malicious Nodes}

\author{Yi Luo, Qiong Li, ~\IEEEmembership{Member,~IEEE}, Hao-Kun Mao    % <-this % stops a space
	\thanks{Yi Luo, Qiong Li and Hao-Kun Mao are with the School of Cyberspace Science, Faculty of Computing, Harbin Institute of Technology, Harbin 150000, China. e-mail: qiongli@hit.edu.cn.}
}
\markboth{Journal of \LaTeX\ Class Files,~Vol.~14, No.~8, August~2021}%
{Shell \MakeLowercase{\textit{et al.}}: A Sample Article Using IEEEtran.cls for IEEE Journals}


\maketitle

\begin{abstract}
%\boldmath
Quantum key distribution (QKD) networks are expected to enable information-theoretical secure (ITS) communication over a large-scale network. Most researches on relay-based QKD network assume that all relays or nodes are completely trustworthy. However, the malicious behavior of any single node can undermine security of QKD networks. Current research on QKD networks primarily addresses passive attacks conducted by malicious nodes such as eavesdropping. 
We suggest a novel paradigm, inspired by distributed systems, to address the active attack by collaborate malicious nodes in QKD networks. Firstly, regarding security, we introduce the ITS distributed authentication scheme, which additionally offers two crucial security properties to QKD networks: identity unforgeability and non-repudiation. Secondly, concerning correctness, our ITS fault-tolerant consensus method, ensures ITS and global consistency with fixed classical broadcast rounds, contrasting with the exponentially message-intensive Byzantine agreement method.  Through our simulation, we have shown that our scheme exhibits a significantly lower growth trend in authentication key consumption compared to the original end-to-end pre-shared keys scheme. 
\end{abstract}


\section{Introduction}
% \fi
\IEEEpeerreviewmaketitle

\IEEEPARstart{Q}{uantum} key distribution (QKD) is a technique that ensures information-theoretic security (ITS) by exploiting the properties of quantum mechanics \cite{Bennett_Brassard_2014}. To expand the application of QKD, numerous researchers have endeavored to construct QKD networks. However, the communication distance in point-to-point QKD (between adjacent nodes) is restricted. Implementing long-range end-to-end (between non-adjacent nodes) QKD necessitates relying on repeaters, such as quantum relays  \cite{Elkouss_Martinez-Mateo_Ciurana_Martin_2013} or trusted relays \cite{Peev_Pacher_Alleaume_Barreiro_Bouda_Boxleitner_Debuisschert_Diamanti_Dianati_2009}) to extend the distance. Due to the challenges in implementing quantum relays  \cite{9684555} (i.e., devices capable of forwarding quantum bits without measurement or cloning) in real-world QKD networks, a more practical approach based on trusted relays has been extensively adopted in prior research \cite{Sharma_Agrawal_Bhatia_Prakash_Mishra_2021,9684555}, with several successful demonstrations of trusted relay-based QKD networks \cite{Wang_Chen_Yin_Li_He_Li_Zhou_Song_Li_Wang_etal._2014,Mao_Wang_Zhao_Wang_Wang_Wang_Zhou_Nie_Chen_Zhao_etal._2018,Sasaki_Fujiwara_Ishizuka_Klaus_Wakui_Takeoka_Miki_Yamashita_Wang_Tanaka_etal._2011,Dynes_2019,Chen_2021,WOS:001009232500020,WOS:000929512600002}

Security is a major concern in QKD research, however, it remains vulnerable to compromise by malicious classical units. The literature \cite{Zapatero_2021} \cite{Curty_2019} highlights that QKD systems implicitly assume trust in classical post-processing units, which is a substantial assumption. Nevertheless, if these classical units have malicious action, the keys security and correctness could be compromised. Although these studies focus on QKD systems, this issue can still be extended to QKD networks.  
%In the QKD network, the point-to-point key distribution task is extended into end-to-end (between non-adjacent nodes) key distribution. 
Numerous studies assumed complete trust in all relays within the network, implying that they would accurately perform the key distribution task without leaking any secret information. However, this assumption is easily undermined in a practical QKD network \cite{Cao_Zhao_Li_Lin_Zhang_Chen_2021,Huang_2022}. For clarity, we define a "malicious node" as any node (or trusted relay) whose classical unit is malicious or untrusted.  Given that these nodes do not need to follow the protocol and can exploit the resources they control to compromise security, addressing malicious behavior becomes pivotal in practical QKD networks.

%Security is a crucial issue in QKD networks, which may be compromised if intermediate trusted relays betray or cracked by attackers.
%%In a trusted relay-based QKD network, trusted relays enable key distribution across multiple QKD systems, which is known as end-to-end key distribution \cite{Zhang_Xu_Chen_Peng_Pan_2018} (QKD between non-adjacent nodes). However, the problem is that when we conduct 
%In end-to-end key distribution (key distribution between non-adjacent nodes), trusted relays (can be referred as trusted node) have access to all the data that is transmitted. If a relay leaks information or is controlled by an attacker, all messages passing through that relay will not be secure any more. In fact, we lack a reliable method for detecting whether a relay has leaked information or is under the control of an attacker. These untrusted relays have the same capability as the normal trusted relays and they are indistinguishable from the information level. 
In recent years, some research efforts have focused on mitigating the potential threats posed by trusted relays.
However, these efforts primarily concentrate on preventing malicious nodes from passive eavesdropping and do not encompass protection against active attacks by malicious nodes. 
Some research proposes using Measurement-Device-Independent QKD (MDI-QKD) \cite{PhysRevLett.108.130503} or Twin-Field QKD (TF-QKD) \cite{Lucamarini_2018} to construct the QKD network, rather than relying on trusted relays \cite{Cao_Zhao_Li_Lin_Zhang_Chen_2021,fan2022robust,xue2022measurement,PhysRevX.6.011024,PhysRevApplied.17.014025}.
These studies propose a hybrid trusted/untrusted network architecture based on MDI-QKD or TF-QKD, which do not need to trust measurement devices. This approach reduces the number of trusted relays, but still requires trusted relays to extend the QKD distance because it does not allow direct connection of two untrusted measurement devices \cite{Cao_Zhao_Li_Lin_Zhang_Chen_2021}.  When a relay that is assumed to be trusted begin to engage in malicious behavior, security is still compromised.   Another approach, such as \cite{ref1}, \cite{Zhou_2022} and
\cite{lo2022distributed}, involve processing key distribution across multiple paths, which can enhance the QKD network's resilience to malicious nodes. It pointed out that we needed to process key distribution on at least $f+1$ disjoint paths to prevent $f$ malicious nodes from eavesdropping information \cite{Zhou_2022}. We found that these research cannot prevent active attacks from malicious nodes proposed in \cite{Zapatero_2021} and \cite{Curty_2019}. Because when the malicious nodes are equipped with their held keys, they can spoof other nodes or disrupt the key distribution process (i.e., disrupting the condition of disjoint paths), presenting a heightened threat.  

% Unfortunately, the solutions proposed by \cite{Zapatero_2021} and \cite{Curty_2019} are not readily applicable to QKD networks. These solutions solely address active malicious devices in point-to-point systems, and their limitation lies in their authentication methods' inability to counter active malicious nodes. \cite{Zapatero_2021} \cite{Curty_2019} employ the authentication scheme using pre-shared keys similar to common QKD systems[]. It is noted in \cite{Wang_2021} that if this scheme were directly applied to QKD networks, it would require pre-sharing $O(N^2)$ pairs of keys. The storage, synchronization, and management of such a large number of key pairs will increase the complexity and security risks of the network. More importantly, the security requirements for point-to-point QKD systems and QKD networks are not entirely the same. When we intend to apply the \cite{Zapatero_2021} and \cite{Curty_2019} scheme to QKD networks, two additional properties need to be considered.
Although the solutions proposed by  \cite{Zapatero_2021} and \cite{Curty_2019} considered active attacks in point-to-point systems, their scheme are not readily applicable to QKD networks. 
In these schemes, trust is established through symmetric point-to-point QKD keys, a method effective only for authenticating adjacent nodes. However, this approach becomes inadequate in identifying non-adjacent nodes, as it necessitates reliance on assistance from other nodes. This reliance introduces a significant security concern, as the assisting nodes themselves could be malicious, exemplifying a fundamental flaw in the point-to-point trust model.

\begin{figure*}[!t]
	\centering  
	\subfigure[Secure multi-path key distribution scenario]{
		\label{fig01}
		\includegraphics[width=0.4\textwidth]{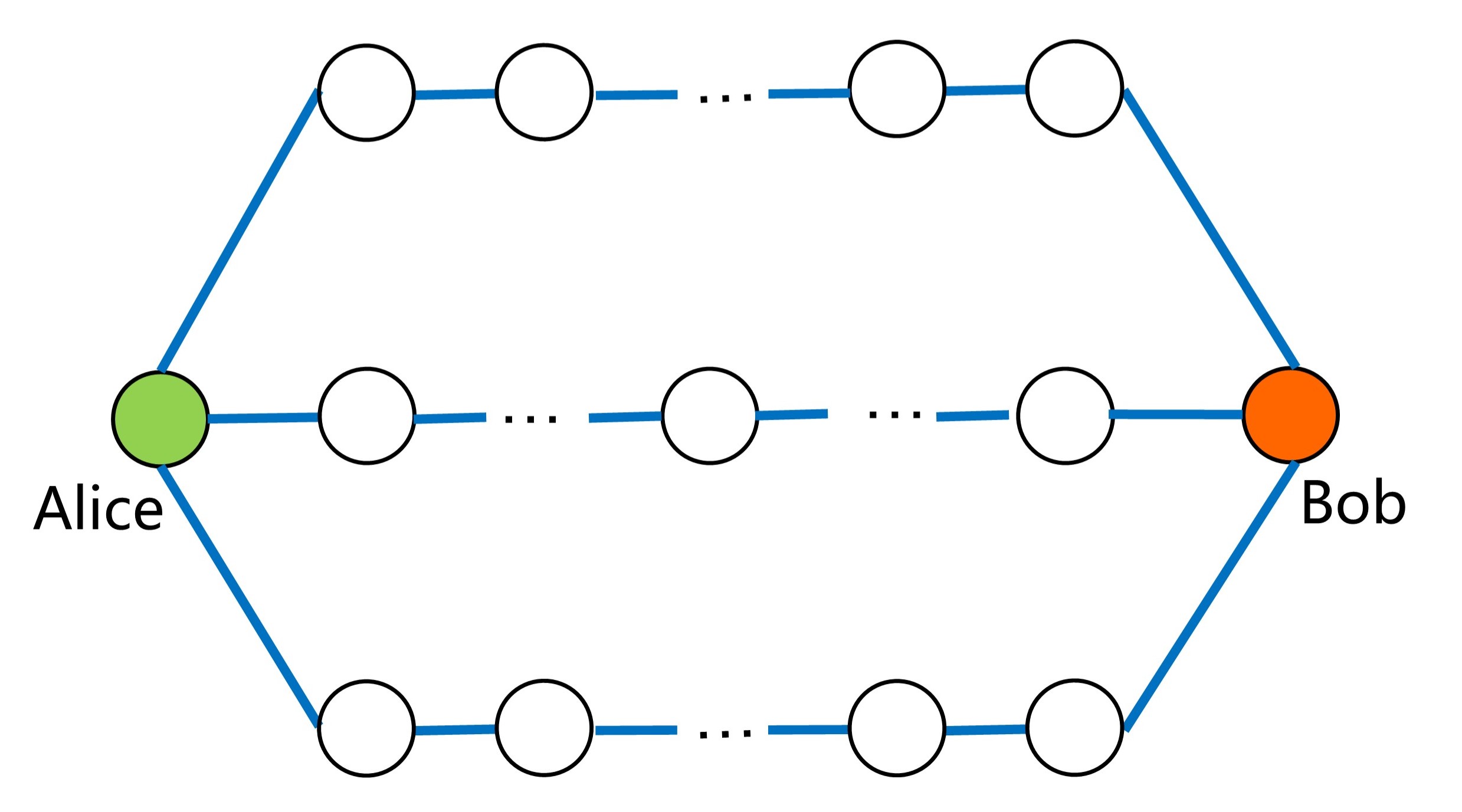}}
	\subfigure[Insecure multi-path key distribution scenario]{
		\label{fig02}
		\includegraphics[width=0.4\textwidth]{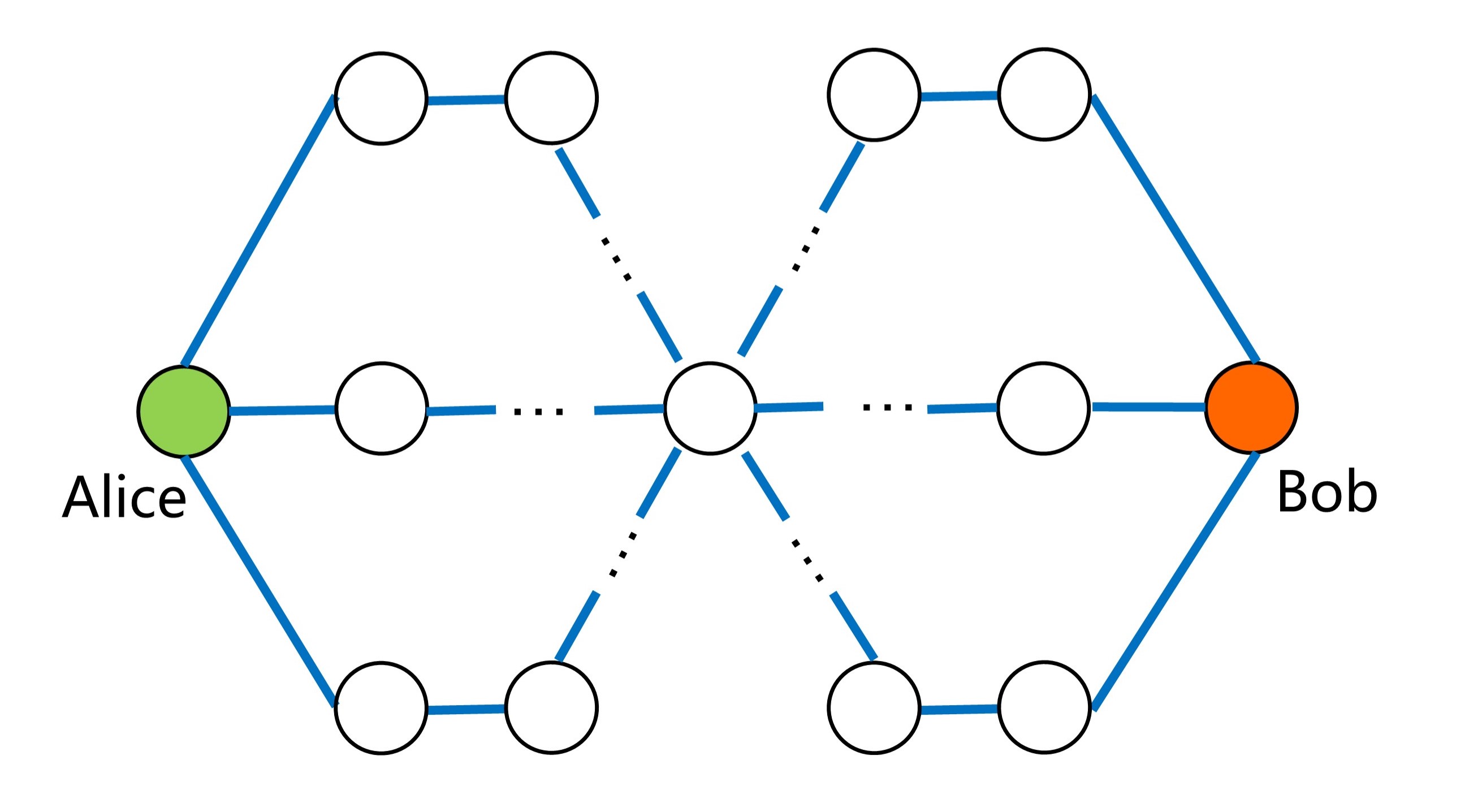}}
	\caption{An example of secure and insecure multi-path key distribution. Assuming that adjacent nodes have already shared symmetric keys using QKD, we now want to use these point-to-point keys to construct an end-to-end symmetric key. The scenario in figure \ref{fig01} is secure because the three paths are non-intersecting, while figure \ref{fig02} may be insecure because the paths have intersections. When the intersection is a malicious node, it can lead to key leakage."}
	\label{fig0}
\end{figure*}

To avoid this threat, we need to consider avoiding the use of cascaded point-to-point modes, and focus on two key properties that have been overlooked in QKD systems.
%These two properties have not been sufficiently considered in QKD systems. However, when extending schemes from QKD systems (such as \cite{Zapatero_2021} and \cite{Curty_2019}) to QKD networks, it is necessary to further consider implementing the following two properties in an information-theoretically secure manner.

1. Identity Unforgeability: 
Even if the attacker Eve collaborates with $f$ malicious nodes, they cannot forge the identity as an honest node or substitute any message from an honest node.  The QKD system involves only two nodes and does not require authentication to distinguish between two nodes. However, within QKD networks, this property is needed to distinguish each node in network. For example, if we use the secret-sharing procedure elucidated in \cite{Zapatero_2021} and \cite{Curty_2019} in QKD networks, this property is needed to ensure that all the secret shares will be send to the correct nodes.

2. Non-repudiation:
Non-repudiation states that any node is not able to refuse authorship of a authenticated message.
Non-repudiation isn't necessary in QKD systems, since with only two nodes, a legitimate message undoubtedly originates from the other node. However, in QKD networks, this property is essential. For instance, if we use the major vote in \cite{Zapatero_2021} and \cite{Curty_2019} in QKD networks, the non-repudiation is necessary to ensures the validity of every votes.

In classical cryptography, these two properties are generally implemented by authentication schemes. Due to the different security assumptions of classical cryptography and QKD assumptions, we cannot directly adopt the methods from classical authentication.
% Achieving the above two properties while ensuring information-theoretic security is challenging.  
 Previous studies employ the authentication scheme using pre-shared symmetric keys similar to common QKD systems.  This method cannot represent a unique identity and therefore does not support the above two properties. Additionally, it is noted in \cite{Wang_2021} that if the pre-shared keys scheme were directly applied to end-to-end nodes, it would require pre-sharing $O(N^2)$ pairs of keys ($N$ represents the number of nodes). The storage, synchronization, and management of such a large number of key pairs will increase the complexity and security risks of the network.

{\bf{Contributions:}} We propose a new paradigm that addresses the aforementioned challenges from both security and correctness perspectives. Our main contributions are summarized as follows:

1. Regarding security, we introduce an ITS distributed authentication scheme to provide identity unforgeability and non-repudiation within the QKD network, which previous preshared-keys authentication cannot provide. Our scheme does not need end-to-end pre-shared keys.  We believe that pre-shared keys is a ideal assumption. Because the pre-shared end-to-end keys may not protected by QKD, we should have doubts about the security of these keys until we achieve end-to-end key distribution in the presence of malicious nodes.  

2. Concerning the issue of correctness discussed in \cite{Zapatero_2021} and \cite{Curty_2019}, we propose a practical ITS fault-tolerance consensus based on our authentication scheme. We have significantly advanced distributed consensus by transitioning from classical security to information-theoretic security. This enhancement guarantees global consistency with a fixed and constant number of classical broadcast rounds. In contrast, as noted by Zapatero (2021), to withstand malicious nodes with ITS, using byzantine agreement requires an exponentially increasing number classical messages among the participating units \cite{Zapatero_2021}

3. Our scheme consumes fewer keys. Through simulation experiments, we found that the authentication keys consumption in the proposed scheme exhibit a smaller growth trend, with authentication key consumption in larger networks (i.e., 80 nodes) being only $13.1\%$ of that in the pre-shared keys scheme.

\section{Problem Statement}
We are considering a scenario where the QKD network is modeled as a graph $G(V,E)$. In contrast to the point-to-point key distribution of QKD systems, the primary concern in QKD networks is the end-to-end key distribution. We assume there are $f$ malicious nodes in QKD network, and their attack capabilities follow the attacker assumptions bellow:

{\bf{Attacker Assumption:}} We assume there are $f$ malicious nodes in QKD networks. Malicious nodes will share all the keys on the QKD links connected to them wich each other. They can utilize these keys not only to passively eavesdropping information but also to actively deceive other nodes by employing their holding keys for authentication. Attackers can exploit authentication keys to intervene in any process of the key distribution, such as the end-to-end key distribution path, and the allocation of key resources on each link for key distribution.

In past methods to prevent malicious nodes, they \cite{Zhou_2022,lo2022distributed} directly utilize the authentication scheme of the QKD system. In this scheme, keys are distributed point-to-point between two adjacent nodes, and then these point-to-point keys are used in a cascaded manner to construct end-to-end authentication keys. During this process, malicious nodes in the middle can actively impact the security of the end-to-end authentication keys. In \cite{ref1}, \cite{Zhou_2022}, and \cite{lo2022distributed}, there is a security condition that the multi-path key distribution requires the paths to be non-intersecting, and the number of keys distributed on each path to be equal. As illustrated in figure \ref{fig0}, we provide an example to distinguish between secure and insecure scenarios of multi-path key distribution. Alice and Bob only possess QKD keys of adjacent nodes and are unable to directly distinguish between the scenarios in figure \ref{fig01} and figure \ref{fig02}. 

Due to the lack of a global perspective, the keys held by Alice and Bob alone are insufficient to determine whether the transmission through multiple nodes in the network (including malicious ones) is still secure. An interesting question is, under the premise of ensuring information-theoretic security, whether it is possible to judge the security and correctness of cooperative key distribution among multiple nodes. To solve this problem, relying solely on point-to-point keys is insufficient, we need a new mechanism to discern the behavior of each node. Our scheme further considers identity unforgeability and non-repudiation. To clearly present our goals, we aim for the proposed scheme to satisfy the following four properties extends from \cite{Zapatero_2021}:

%Key distribution in QKD networks is more complex compared to QKD systems, the variables related to the security of key distribution will increase. In QKD networks, multiple end-to-end key distributions may occur simultaneously.  Each node can act both as a source node proposing key distribution requests and as a relay assisting other nodes. They can determine the path of the key distribution process and the amount of keys consumed on each QKD link. 

%Attackers can also completely ignore the message propagation time, either not sending messages according to the protocol schedule or delaying the transmission of some messages.

{\bf{Security: }}For each node, achieving identity unforgeability and non-repudiation. During the process of key distribution, the mutual information between the final key generated by the source and destination nodes and the keys held by other relay nodes can be neglected.

{\bf{Consistency: }} All honest nodes agree on the specifics of the multi-path key distribution,  including the path that the key distribution will take and the amount of keys consumed on each edge.

{\bf{Conditional correctness: }} Assuming an honest source node intends to distribute a key $s$, if consistency can be satisfied, all honest destination nodes output $s$.

{\bf{Fairness: }} The key distribution requests from each node are finished with equal probability.  Malicious nodes will not gain any additional advantages.

\section{Proposed Distributed Information-theoretical Secure Protocols}\label{sec2}

\subsection{\label{authentication}Overview}

We begin by providing an overview of our scheme. The goal of our approach is to achieve information-theoretic secure key distribution, even in the presence of a collaboration of $f$ malicious nodes.  Our scheme can accommodate up to $ MIN\left( C-1,\lfloor \frac{N-1}{2} \rfloor \right)  $ malicious nodes, where $C$ is the node connectivity of the network and $N$ is the number of nodes in the network.  $\lfloor~\rfloor$ represents the floor function.

%We will analyze this upper bound in more detail in security analysis section.

In our scheme, messages with propagation delay exceeding $ \varDelta $ will be discarded, where $ \varDelta  $ is a known upper bound on the propagation delay \cite{Abraham2020Sync}.  Fortunately, determining whether a message has timed out is simple because each QKD link includes a time synchronization mechanism in the physical layer \cite{Wang_Chen_Guo_Yin_Li_Zhou_Guo_Han_2012,Pljonkin_Rumyantsev_Singh_2017}, which is more accurate than traditional network time synchronization. Any attacker interfering with time synchronization will result in a variation quantum bit error rate. Any nodes can determine the synchronization status by cross-referencing all the QKD links. If the maximum synchronization error exceeds the setup threshold, then this node temporarily abandon the current protocol.

%Malicious nodes may not adhere to this rule, and even the time synchronization between malicious nodes and other nodes can be incorrect.
%In addition, since the connectivity $C$ of the network exceeds the number of malicious nodes $f$, there are at least $f+1$ disjoint paths between any two nodes and the nodes can determine the synchronization status by cross-referencing all paths. In addition, this time synchronization in QKD physical layer, which has not been fully exploited in previous studies, offers the possibility of implementing a more security authentication scheme without a pre-shared key. 
%We further illustrate this point with examples in the appendix A.

\begin{figure*}[!t]
\centering
\includegraphics[width=5in]{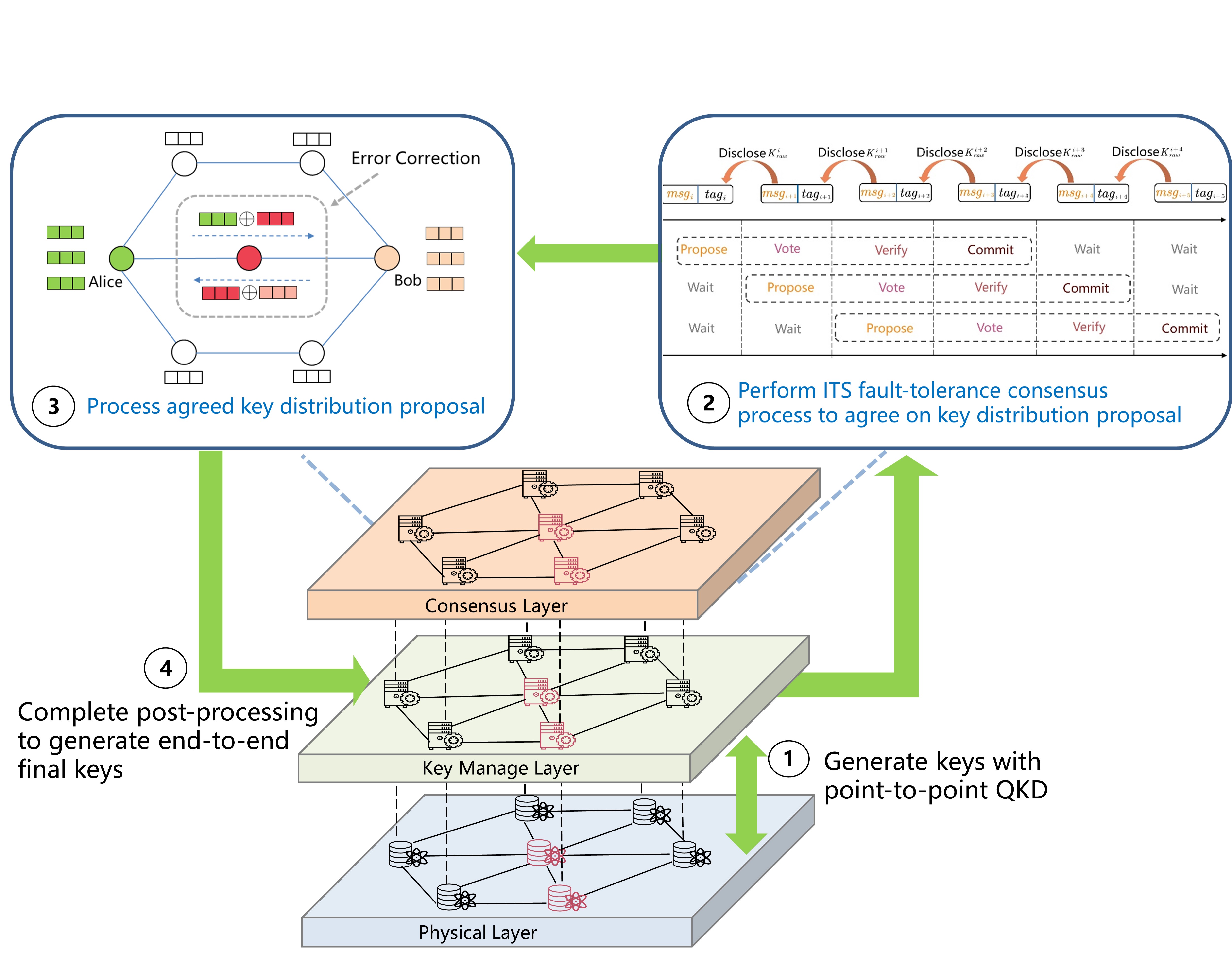}
\caption{Overview of the proposed architecture. There are four main steps here. The first and second steps can be referred to existing QKD schemes. We will provide all the details for the second and third steps in section \ref{BFT} and \ref{Execution}.}
\label{fig1}
\end{figure*}

{\bf{Architecture:}}
The architecture of our scheme is shown in the figure \ref{fig1}, there are four steps to complete the key distribution process in our scheme.
\begin{enumerate}
\item{Each adjacent node generates QKD keys via a point-to-point QKD link between them.}
\item{Each node participates in the consensus process and agrees on a key distribution proposal. The key distribution proposal includes the distribution paths required to implement key distribution and the number of keys allocated to each path. The purpose of consensus is to achieve agreement among all honest nodes, ensuring that participating nodes cooperate correctly and avoid being misled by malicious nodes.}
\item{Each node processes the key distribution proposal which is agreed in consensus.}
\item{ Source and destination nodes of the key distribution proposal perform post-processing (checking correctness and privacy amplification) to generate the final end-to-end keys.}
\end{enumerate}

Steps 1 and 4 can be referred to existing QKD schemes. We will focus on step 2 and 3 in the following. Here we first introduce the ITS distributed authentication in \ref{authentication}, as it will be used in messages propagation. The workflow of consensus process and the processing of key distribution proposal will be described in \ref{BFT} and \ref{Execution}.

\subsection{\label{authentication}ITS Distributed Authentication }
The ITS distributed authentication scheme consists of four functions: authentication key generation, authentication tag generation, authentication key disclosed and authentication verification. Here, we divide the timeline into several time intervals based on the upper limit of message propagation $ \varDelta  $. The starting point of the $i$th interval is recorded as $T_i$.

%It is important to note that BS has a valid time $\varDelta$ and BS received outside of that time should be discarded. 

\begin{figure}[!t]
\centering
\includegraphics[width=3.5in]{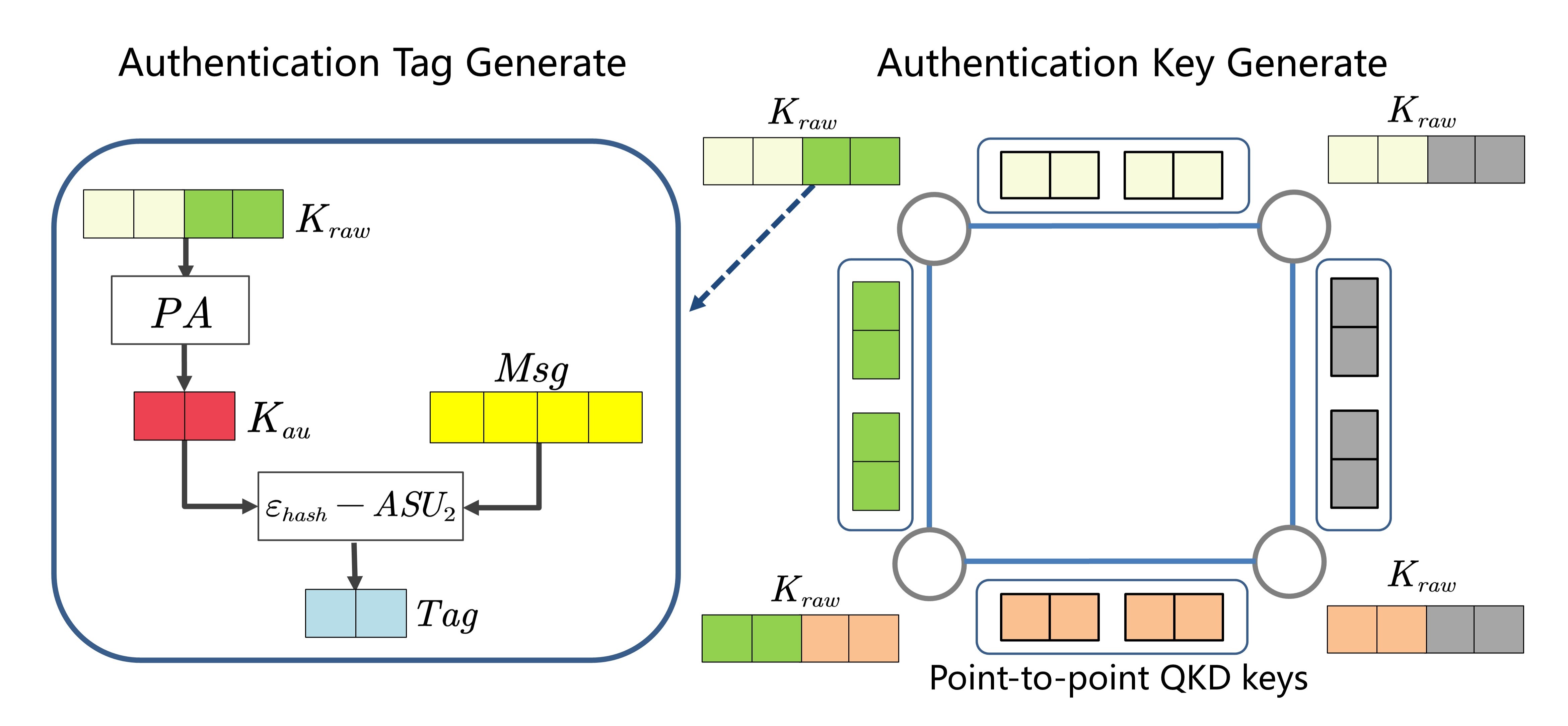}
\caption{ Authentication key generation and authentication tag generation process.}
\label{fig3}
\end{figure}

{\bf{1. Authentication key generation:}}  For any given node $v \in V$, the input to the key generation function consists of all point-to-point keys on the QKD links that are connected to node $V$ and the output is the authentication key $K_{au}$. As shown in figure \ref{fig3},  assuming that node $v$ has $x$ adjacent edges, node $v$ extracts point-to-point QKD keys $K_{adj_1}$ from each adjacent QKD link and concatenate them to $ K_{raw} $,  which is denoted as \eqref{eq2}.  Here $K_{adj_1},K_{adj_2}...K_{adj_x}$ have the same lenght $k$, which is calculated by \eqref{eq0}. The $\varepsilon _{au}$ is security parameter, $C$ is the node connectivity of QKD networks, $|K_{au}| $ is the length of the $K_{au} $, $s$ is the parameter of the privacy amplification and $f$ is the number of malicious nodes. We will analyze equation \eqref{eq0} in detail in security analysis.
\begin{equation}
\label{eq2}
K_{raw}=K_{adj_1}||K_{adj_2}...||K_{adj_x} 
\end{equation}
\begin{equation}
\label{eq0}
k=MAX\left( \frac{-\log _2\varepsilon _{au}}{C-f},\frac{|K_{au}|+s}{x-f} \right)  
\end{equation}

After generating $K_{raw}$, node $v$ use privacy amplification hash function $G$ to compress $K_{raw}$ into $K_{au}$, which is expressed in \eqref{eq3}.
\begin{equation}
\label{eq3}
K_{au}=PA\left(G, K_{raw} \right)  
\end{equation}

{\bf{2. Authentication tag generation:}} At time $ T_i $, we assume $msg_i$ is the message (with timestamp) needed to be authenticated. Then, we use $K_{au}$ choose a hash function $h_{K_{au}}$ in a $ \varepsilon _{hash}-ASU_2  $ family to generate authentication tag $tag_{i}$, denoted as \eqref{eqq3}. The $ \varepsilon _{hash}-ASU_2  $ can be constructed by many ways, such as \cite{Kiktenko_Malyshev_Gavreev_Bozhedarov_Pozhar_Anufriev_Fedorov_2020, Abidin_2012}. The authentication tag is just valid for $ T_i-T_i+\varDelta  $ time. The  $tag_i$ and  $msg_i$ will be broadcast to other nodes in the QKD networks.
\begin{equation}
\label{eqq3}
tag_{i}=h_{K_{au}}\left( msg_i \right) 
\end{equation}

{\bf{3.Authentication key disclosed:}} Assuming no attacker interference with message propagation, a message and its authentication tag from node $v$ at time $ T_i $ will reach other nodes before $T_i+\varDelta$. Nodes first check the message's timestamp upon receipt. If the timestamp deviates by more than $\varDelta$ from a node's local time, the message is discarded. Otherwise, it's retained for verification. If node $v$ broadcasts a message at time $ T_i $, then at $T_i+\varDelta$, it will disclose the key $K_{raw}^i$ to assist other nodes in completing the authentication process.

{\bf{4.Authentication verification:}}  A node will receive broadcasts from multiple other nodes, including $ tag_i $ and $msg_i$ sent at time $ T_i $ as well as $K_{raw}^i$ sent at time $T_i+\varDelta$. We denote the content from the same node as a tuple $(K_{raw}^i, msg_i, tag_i )$. Authentication verification of the  requires it to pass through the following two steps:

\begin{itemize}
\item  First, regenerate the tag using $K_{raw}^i$  and $msg_i$ according to the authentication tag generation process, and then determine whether it is equal to $ tag_i $. After check all received tuples, verify whether among these correct tuples, there are consistent keys of length $k$ between $ K_{raw}^i $ of neighboring nodes. If there are consistent keys of length $k$ between $ K_{raw}^i $ from two nodes, a edge will be recorded between these two nodes. After all $ K_{raw}^i $ are checked, a graph will be formed and referred to as the validation key graph.
%	\begin{equation}
	%		\label{eq22}
	%		\begin{aligned}
		%				K_{au}^i=PA\left( K_{raw}^i \right) \\
		%			h_{K_{au}^i}\left( m_i \right)  = t_{i}
		%		\end{aligned}
	%	\end{equation}
\item  Second, each node consults the validation key graph to establish its trust relationship with other nodes. If there is at least $f+1$ path in the validation key graph from the source node to the destination node, then the message broadcasted by the source node can be trusted by the destination node.
\end{itemize}

Our scheme is a one-time authentication scheme that primarily leverages the information asymmetry on time between the source node of message and other nodes.
Before Authentication key disclosed step, only the source node possesses knowledge of the authentication key, while other nodes and attackers know nothing about it. We have set a limit on the maximum message transmission delay $ \varDelta  $, and messages that exceed this time are discarded. Attackers may attempt to modify messages during the message propagation process, but without the authentication key, they cannot ensure that the modified messages will pass authentication. Under normal circumstances, before time $ \varDelta  $, the message will be accepted by the receivers and locked. The message source node only discloses the authentication key after a certain time delay $ \varDelta  $ from sending the message. This ensures that even if an attacker witnesses the key disclosure, they cannot further modify the messages stored by the receivers.

\subsection{ \label{BFT} ITS Fault-tolerance Consensus Workflow}

Our ITS distributed authentication scheme just addresses the issue of ensuring secure message propagation in the presence of malicious nodes. To achieve the correctness and consistency of end-to-end key distribution in the presence of malicious nodes, we propose the ITS fault-tolerant consensus. 

Traditional consensus schemes rely on classical cryptography signature, whereas our consensus scheme relies on the keys provided by QKD, making it information-theoretically secure. Our ITS distributed authentication scheme can be integrated with consensus mechanisms. We have provided an implementation with the synchronization consensus approach from \cite{Abraham2020Sync} and \cite{abraham_et_al:LIPIcs.OPODIS.2021.27}. The strength of this scheme lies in its capability to reach consensus for multiple different key distribution processes within a constant number of broadcast rounds. Furthermore, under synchronous conditions, this consensus scheme can handle up to $\lfloor \frac{N-1}{2} \rfloor$ malicious nodes, as evidenced in \cite{Abraham2020Sync, abraham_et_al:LIPIcs.OPODIS.2021.27}.

We combine our proposed ITS distributed authentication scheme with the synchronization consensus approach from \cite{Abraham2020Sync} and \cite{abraham_et_al:LIPIcs.OPODIS.2021.27}. The advantage of this scheme lies in its ability to achieve consensus for multiple different key distribution processes within a constant number rounds broadcast. 
Additionally, under synchronous conditions, the consensus scheme can accommodate a number of malicious nodes up to $\lfloor \frac{N-1}{2}  \rfloor$, which has been proven in \cite{Abraham2020Sync, abraham_et_al:LIPIcs.OPODIS.2021.27}.
%In the broadcast at time $T_i+\varDelta$, disclosing the authentication key from the previous broadcasting at $T_i$. This method allows us to achieve authentication in a pipelined manner within the consensus protocol. We will provide a detailed description in section \ref{BFT}.

%
%Our solution further takes into account the possibility of attackers manipulating message delays, which is referred to as the mobile sluggish model \cite{10.1145/357172.357176} in classical networks. In this scenario, the consensus solution requires that $f$ satisfies the $f \le \lfloor \frac{N-1}{2}  \rfloor$.  
% our solution can accommodate $ MIN\left( C-1,\lfloor \frac{N-1}{2} \rfloor \right)  $ malicious nodes. 

In this section, we detail the workflow of the proposed ITS fault-tolerance consensus.  In our consensus scheme, there are two roles: leader and replica. The leader is responsible for proposing the key distribution proposal, and the leader's term is identified by a view number. Replica is responsible for voting and passing legitimate proposals if possible. 

%To ensure fairness, there is a upper limit $\gamma$ of distribution keys in each view and each view changes its leader according to the BFT signature keys disclosed in the previous view. 

\begin{figure}[!t]
\centering
\includegraphics[width=3.5in]{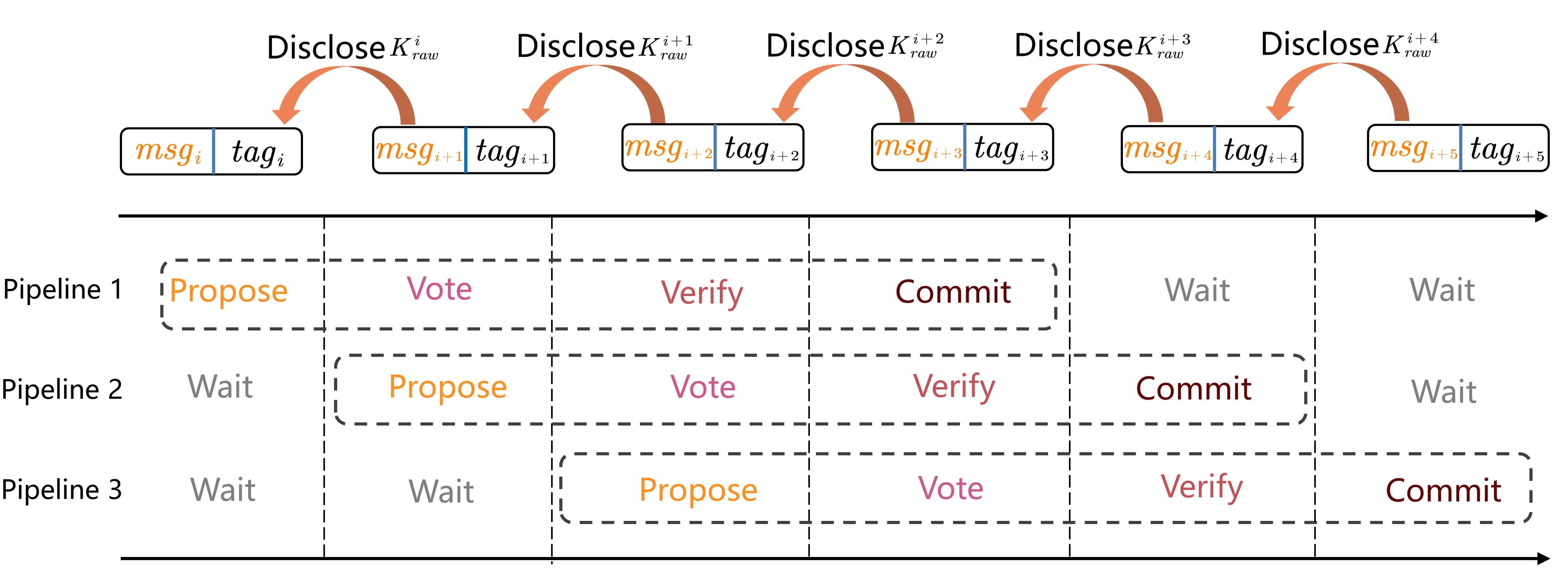}
\caption{Workflow and key schedule of ITS fault-tolerance Consensus. There are three identical pipelines, and each pipeline executes the same ITS fault-tolerance consensus workflow. During broadcasting, messages from all three pipelines can be merged and broadcast together. The broadcasting occurs at regular intervals of $\varDelta $, and with each broadcast, the authentication key from the previous broadcast is disclosed.}
\label{fig2}
\end{figure}

As shown in figure \ref{fig2}, a view includes 4 steps, in which all replicas broadcast a message to participate in the consensus. 
Any message broadcast at time $T_i$ will be verified by the authentication key disclosed at time $T_i+\varDelta  $, forming an overall cascaded verification process.
%The BFT signatures of the broadcasted messages are generated using different BS keys, and messages broadcasted at time $T_i$ are verified with the BS keys disclosed at time $T_i+\varDelta  $.

{\bf{Message Format: }}Messages are broadcast as $ \left< SN, \left<msg_i\right>, e, K_{raw}^{i-1},tag_i \right> _{Source} $. $SN$ represents the unique message serial number, $m_i$ represents the message to be broadcast at time $T_i$, $e$ represents the view number which the message belongs, $tag_i$ is the authentication tag of the content to be broadcast. $K_{raw}^{i-1}$ is the disclosed key to verify the $T_{i-1}$ authentication tag. $source\in \left\{ l,r,a \right\}$ represents the node who broadcast the message, $l$ represents the leader, $r$ represents replicas and $a$ represents all nodes.

{\bf{Legitimate proposal: }} The proposal can be viewed as a multi-path key distribution scheme. In the proposal, it is necessary to specify the nodes through which each path passes (or the nodes participating in secret sharing) and the parameters, such as the amount of the keys on each path.
Multiple proposals regarding different source nodes and destination nodes can be bundled together into $P_e$. Any node can verify whether a proposal is legitimate.

\begin{figure}[!t]
	\centering
	\includegraphics[width=3.5in]{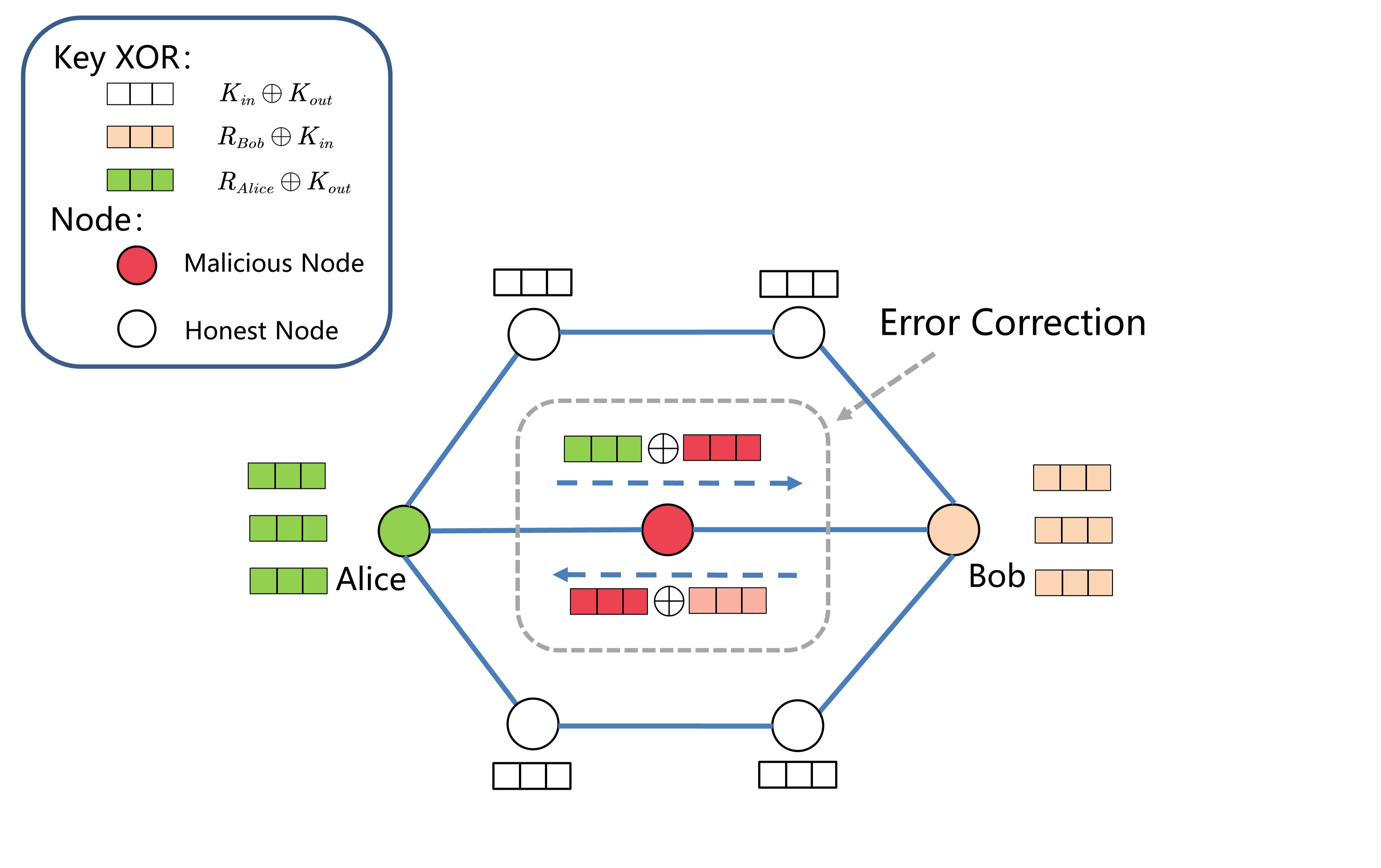}
	\caption{An illustration of the processing of end-to-end key distribution. For any node on a given path, the key on the incoming link to that node is $K_{in}$, and the key on the outgoing link from that node is $K_{out}$. $R_{Alice}$ and $R_{Bob}$ are respectively the random numbers chosen by Alice and Bob on each path.}
	\label{fig4}
\end{figure}

{\bf{Multi-path key distribution: }} The multi-path key distribution scheme can be represented as a subgraph $G_p$. We provide a simple example based on a multi-path key distribution scheme. In multi-path key distribution scheme, the number of disjoint key distribution paths need to greater than $f$ and the amount of keys distributed on each path should be equal.  Figure \ref{fig4} illustrates the case when $f=2$. The source node for key distribution is Alice, and the destination node is Bob. There are three disjoint paths connecting them. 
Assuming that $S$ is the key Alice wishes to distribute, Alice will generate a random number $r$ for each path.
In our example, there are three random numbers, $R_{A1}, R_{A2}, R_{A3}$, which satisfy $R_{A3} = S \oplus R_{A1} \oplus R_{A2}$. Alice will also generate syndromes $sd(R_{A1}),sd(R_{A2}),sd(R_{A3})$ to verify $R_{A1}, R_{A2}, R_{A3}$ later.  
The proposal will contain the multi-path key distribution scheme $G_p$, the lengths of key distribution for each path, and the syndromes $sd(R_{A1}),sd(R_{A2}),sd(R_{A3})$.

{\bf{Key XOR:}} Key XOR (KX) are the Xor results of the keys on the input and output links of a node in the key distribution scheme $G_p$.

{\bf{Initial configuration: }}To avoid idle waiting, our consensus protocol has established three pipelines. The protocol workflow within each pipeline is identical. Here, we will only describe the protocol workflow within one pipeline. 
During the initial process, it needs to set the number of malicious nodes $f$ and the maximum transmission time $\varDelta $. While the consensus protocol is in progress, a timer of length $ 8\varDelta  $ is set for each leader. The workflow of ITS fault-tolerance consensus is described as follows:

\begin{enumerate}
\item{{\bf{Propose: }} When a node is chosen as the leader, it needs to form a legitimate proposal and then broadcast $ \left< Propose, \left<P_e\right>, e, K_{raw}^{i-1}, tag_i \right> _l $. }
\item{{\bf{Vote: }}When a replica receives the $Propose$ message, if the key distribution proposal $P_e$ is legitimate, it can vote on the proposal. The replica will calculate the syndrome of $KX$ based on the content of the proposal. If the proposal does not require the involvement of that node, this item remains empty. Replica broadcast the voting message with the syndrome of $KX$, proposal $P_e$ and its authentication tag $tag_i$, denoted as $ \left< Vote,\left< sd(KX), P_e,  tag_i \right>, e, K_{raw}^{i}, tag_{i+1} \right> _r $. }
\item{{\bf{Verify: }} Each node broadcasts a verification message $\left< Verify, \left<SN\right>, e, K_{raw}^{i+1} , tag_{i+2} \right> _a $ to verify the last message. $SN$ is the serial number of the message to be verified. }

%	\item{{\bf{Pre-commit: }} 
	%		Each node checks whether the received propose message of the leader is valid and whether there are equivocate proposal. If two equivocate $P_{e+1}$ are found, then a view change message is broadcast with the equivocate proposal. If there are no equivocate $P_{e+1}$, each node will check the number of valid votes for $P_{e+1}$.  If all nodes involved in key distribution as mentioned in the proposal have broadcast their votes, and the total number of votes is greater than $f$, the replica broadcast $\left< Commit, \left<P_{e}, sd(KX)_{set}\right> , e, K_{raw}^{i+2}, tag_{i+3} \right> _a $. The $sd(KX)_set$  represents the set of $sd(KX)$ from all valid votes.} 
%	%		If a node finds that the $ P_e$ it voted for is incorrect, the node can change its vote. Otherwise it maintains its original vote. Each node broadcasts a new vote $ \left< Revote,\left< P_e, Propose_{BS} \right>, e, BS\left( K_{e}^{3} \right), K_{e}^{2} \right> _a $. }
%
%
%\item{{\bf{Commit verify: }}  Each node broadcasts a verification message to verify the $Propose$ and $Vote$ $ \left< Verify_C, \left<SN\right>, e, K_{raw}^{i+3}, tag_{i+4} \right> _a $. $ SN$ is the serial number of the message to be verified. }
\item{ {\bf{Commit: }} Each node checks whether the received propose message of the leader is valid and whether there are equivocate proposal. If two equivocate $P_{e}$ are found, then a view change message is broadcast with the equivocate proposal. If there are no equivocate $P_{e}$, each node will check the number of valid votes for $P_{e}$.  If all nodes involved in key distribution as mentioned in the proposal have broadcast their votes, and the total number of votes is greater than $f$, set commit-timere to $2\varDelta$ and start counting down. When commit-timer reaches 0, if no view change has been detected, commit $P_{e}$ .
}

\end{enumerate}

{\bf{View change: }}  The view change message is triggered by the following situations:
\begin{enumerate}
\item{ Any node finds that leader proposes two equivocate proposals.}
\item{ Leader fails to commit a proposal within the set time.}
\item{ Failing during the verification process.}
\end{enumerate}	
When a node discovers the conditions for view change, it needs to broadcast view change messages at each step until timeout or it receives $f+1$ view change messages with valid signatures. The format of the view change message is  $ \left< View change,\left< Event \right>, e,  K_{raw}^{i} , tag_{i+1} \right> _r $. $Event$ represents the event that triggers view change.

{\bf{Fairness leader election: }} 
The leader is replaced under the following two conditions:
\begin{enumerate}
\item{On receiving $f + 1$ valid votes for $P_{e}$.}
\item{On receiving $f + 1$ viewchange message.}
\end{enumerate}	
Here view change is a special case. Upon entering new epoch $e$ by view change, if new leader has last $P_{e-1}$, it proposes immediately; otherwise, it waits for $2\varDelta$ time to ensure it can receive the newest $P_{e-1}$ from honest replicas.

The computation of the next leader depends on a common random string $CRS$. The $CRS$ (Common Random String) originates from the last proposal $P_{e-1}$ that achieved consensus. Proposal $P_{e-1}$ contains a multipath key distribution proposal. If this proposal can be committed, then each node participating in the multipath key distribution has voted it in epoch $e-1$. The set of these nodes is denoted as $ V_{e-1}$. Since the multipath key distribution scheme requires $ f+1 $ paths, $ |V_{e-1}|$ is greater than $ f+1 $. Assuming the authentication key of the voting message of node $v \in V_{e-1} $ is $K_{raw}^{v}$, then the CRS is derived using equation \ref{eq1}. Since $K_{raw}^{v}$ is generate from QKD and $|V_{e-1}|$ is greater than $ f $,  $f$ malicious nodes cannot control the next leader. The probability of each node becoming the leader is almost entirely equal. 

%The leader election relies on the disclosed key $ Key_v $ of each node, which can be seen as a random number $ R_v $.
%$v$ is the source of the disclosed key. In our proposed scheme, whether a node can eventually become a leader is unrelated to its $ R_v $. To participate in the leader election, the currently agreed disclosed keys are used as the random number. The quantum random number of each node is noted as $ R_i $, where $i$ is the label of the node. In the leader election process, each node's votes can be calculated with the formula \eqref{eq1}. The next leader is the node $i$ that its $Vote_i$ has the minimal value.
\begin{equation}
\label{eq1}
CRS = \bigoplus_{v\in V_{e-1}}^{}{K_{raw}^{v}}
\end{equation}

\subsection{\label{Execution} Process Agreed Key Distribution Proposal}
End-to-end key distribution will be process after each node has reached consensus on the key distribution proposal $P_{e}$. If malicious nodes act as relays and directly participate in the encryption and decryption process of key distribution, they can still potentially compromise the correctness. Therefore, our protocol requires relay nodes to send the key materials needed for key distribution directly to the source and destination nodes. As shown in figure \ref{fig4}, the key distribution will be accomplished with the assistance of Key XOR.

1. When a node finds itself specified as an intermediate node for key distribution in the proposal, it will generate Key XOR according to the instructions in the proposal. It then sends Key XOR to the source and destination nodes specified in the proposal. Since the syndromes of all the KX has been broadcast to all nodes during the voting process, it can directly check the correctness of KX through the results of consensus.

2. When the source node Alice receives Key XOR from all the nodes on $f+1$ paths. Alice XORs the Key XOR from all the nodes on each path, and sends the result to the destination node Bob. Bob can verify the correctness of each path by checking the syndromes of the individual path keys mentioned in the proposal.

3.After receiving $f+1$ results from different paths, Bob can verify the correctness of each path by checking the syndromes of the individual path keys mentioned in the proposal. If they are all correct, use the PA to compress their results into the final end to end key.
%{\bf{Key XOR Transmission:}}
%Key XOR Transmission (KXT) means to perform Xor operations on all KXs on a key distribution path except the source node and destination node.

Our scheme makes maximum efforts to ensure correct key distribution. Under the condition of successful consensus, $f+1$ syndromes guarantees that the KX received by the source and destination nodes will not be inconsistent. If any malicious node intentionally broadcasts the wrong KX, the error can be eliminated by XOR KX twice in both directions. The key on this path is considered to be exposed to malicious nodes and the attacker can be traced later based on the authentication tag.

\section{Security Analysis}
In this section, we will address the two properties mentioned in the introduction. We will also analyze the maximum number of malicious nodes that the network can accommodate and provide an explanation for why the number of key extractions from QKD links needs to satisfy Equation \ref{eq0}.
%This section focuses on analyzing security properties of our scheme. We will analyze the information-theoretical security of key delivery, information-theoretical security of BFT signature and consensus security. We prove that the ITS of key delivery and BFT signature requires $f$ to be at most $C-1$, and consensus security requires $f$ to be at most $\lfloor \frac{N-1}{3} \rfloor$.

\subsection{\label{B1} Identity Unforgeability}
Identity unforgeability necessitates the consideration of two scenarios \cite{shiu_2007}: substitution attacks and impersonation attacks. Substitution attacks entail an attacker replacing an existing identity, whereas impersonation attacks involve the attacker introducing a new identity. We will address these in two distinct sections.

{\bf{1. Substitution Attacks:}}

If an attacker intends to replace the identity of an honest node, the most direct method would forge the authentication key associated with that node. Here we analyze the mutual information between the attacker's union $E$ and the authenication key $ K_{au} $ and prove that attacker's mutual information about $ K_{au} $ satisfies \eqref{eq7}. We assume that $ K_{raw} $ comes from $x$ neighboring QKD links. We note that the length of $ K_{raw} $ is $n$, the length of $ K_{au} $ is $m$, the secret parameter of PA is $\varepsilon _{PA}$.

%	\label{eq5}
%	k=\frac{n-s-m}{x} 
%\end{equation}

%\begin{equation}
%	\label{eq6}
%	K_{au}=G\left( K_{raw} \right)   
%\end{equation}

\begin{equation}
\label{eq7}
I_{KE}=I\left( K_{au}:E\left( K_{raw} \right) ,G \right) \le \ln \left( 2^{m+fk-n}+ 2^m\varepsilon _{PA} \right) /\ln 2   
\end{equation}

\begin{proof}

According to information theory, $ I_{KE}  $ holds that \cite{Bingze_2021}:

\begin{equation}
\label{eq8}
\begin{aligned}
	I_{KE}  &=H\left( K_{au} \right) -H\left( K_{au}|E\left( K_{raw} \right) ,G \right)  \\  
	&\le H\left( K_{au} \right) -H_{\text{Re}n}\left( K_{au}|E\left( K_{raw} \right) ,G \right)  
\end{aligned}
\end{equation}

$ H_{\text{Re}n}\left( \cdot \right)  $ is Renyi entropy function, which is defined on the probability space $ \left( X,P\left( x \right) \right)  $. 

\begin{equation}
\label{eq9}
\begin{aligned}
	H_{\text{Re}n}\left( X \right) &=-\log _2\varDelta p_x \\
	\varDelta p_x &=\sum_{x\in X}^{}{\left( P\left( x \right) \right) ^2} 
\end{aligned}
\end{equation}

The Renyi entropy is proved to satisfy the lemma 1 in \cite{Stinson2010UniversalHF}.

\begin{lemma}
Let $ r\in R $ denote a variable chosen randomly and $ w\in W $ denote the output of eavesdropping $ w=E\left( r \right)  $. $G$ is a universal hash function chosen from $ \varepsilon_{PA}-ASU_2 $ hash family. $Y$ is a random variable with respect to  $ y=G\left( r \right)  $. then the following equation holds
\begin{equation}
	\label{eq13}
	H_{\text{Re}n}\left( Y|w,G \right) \ge -\log _2\left( \varDelta p_{r|w}+\varepsilon \right)
\end{equation}
\end{lemma}

When Eve controls $f$ nodes around the node which process authentication, $|W|=|E\left( K_{raw} \right) |=2^{fk} $, $|R|=2^n$ and $\varDelta p_{r|w} = \frac{|W|}{|R|}$. We can derive equation \eqref{eq10} from lemma 1.

\begin{equation}
\label{eq10}
H_{\text{Re}n}\left( K_{au}|E\left( K_{raw} \right) ,G \right) \ge -\log _2\left( \frac{\left| W \right|}{\left| R \right|}+\varepsilon _{PA} \right)    
\end{equation}

According to equation \eqref{eq8} the mutual information holds that:
\begin{equation}
\label{eq11}
\begin{aligned}
	I_{KE}  &\le H\left( K_{au} \right) -H_{\text{Re}n}\left( K_{au}|E\left( K_{raw} \right) ,G \right) \\
	&\le H\left( K_{au} \right) +\log _2\left( \frac{\left| E\left( K_{raw} \right) \right|}{|K_{raw}|}+\varepsilon _{PA} \right) \\ 
	&\le \log _2\left( 2^{H\left( K_{au} \right)}\cdot\frac{\left| E\left( K_{raw} \right) \right|}{|K_{raw}|}+2^{H\left( K_{au} \right)}\cdot\varepsilon _{PA} \right)\\ 
	&\le \ln \left( 2^{m+fk-n}+2^m\varepsilon _{PA} \right) /\ln 2  
\end{aligned}
\end{equation}
\end{proof}

A special case arises when we use Toeplize-based PA, \cite{Kiktenko_Malyshev_Gavreev_Bozhedarov_Pozhar_Anufriev_Fedorov_2020} point out that $ \varepsilon _{PA}=1/2^m $. Since $ \ln \left( x+1 \right) \le x $, it can be derived that:
\begin{equation}
\label{eq12}
I\left( K_{au}:E\left( K_{raw} \right) ,G \right) \le 2^{m+fk-n}/\ln 2 
\end{equation}

If we denote $ s=n-fk-m $, then we can get $ I\left( K_{au}:E\left( K_{raw} \right) ,G \right) \le 2^{-s}/\ln 2 $.  If we choose the proper safety parameters $s$, 
the attacker has almost no information about the authentication keys.

{\bf{\label{B4}2. Impersonation Attacks:}}

Another type of attack is for Eve to use her known key to impersonate a non-existent fake node and then make a fake vote to gain an advantage during the consensus process. 
The attackerd creates a fake tuple $(K_{raw}', G', K_{au}', msg', tag')$ in an attempt to deceive other nodes during verification. For the attackerd to successfully execute this attack, they must ensure that the $K_{raw}'$ they choose have an intersection with the keys of the neighboring nodes. For example, at time $T_i$, the attacker needs to determine $(G', K_{au}', msg', tag')$. Then, at time $T+\varDelta$, he discloses $K_{raw}'$. The $K_{raw}'$ need to have a length $k$ intersection with the $K_{raw}$ of neighboring nodes. Firstly, $k$ needs to satisfy equation \eqref{eq27} to ensure there is enough key material to generate $K_{au}$.
\begin{equation}
\label{eq27}
k(x-f) \ge |K_{au}|+s 
\end{equation}

\begin{figure}[!t]
\centering
\includegraphics[width=3.5in]{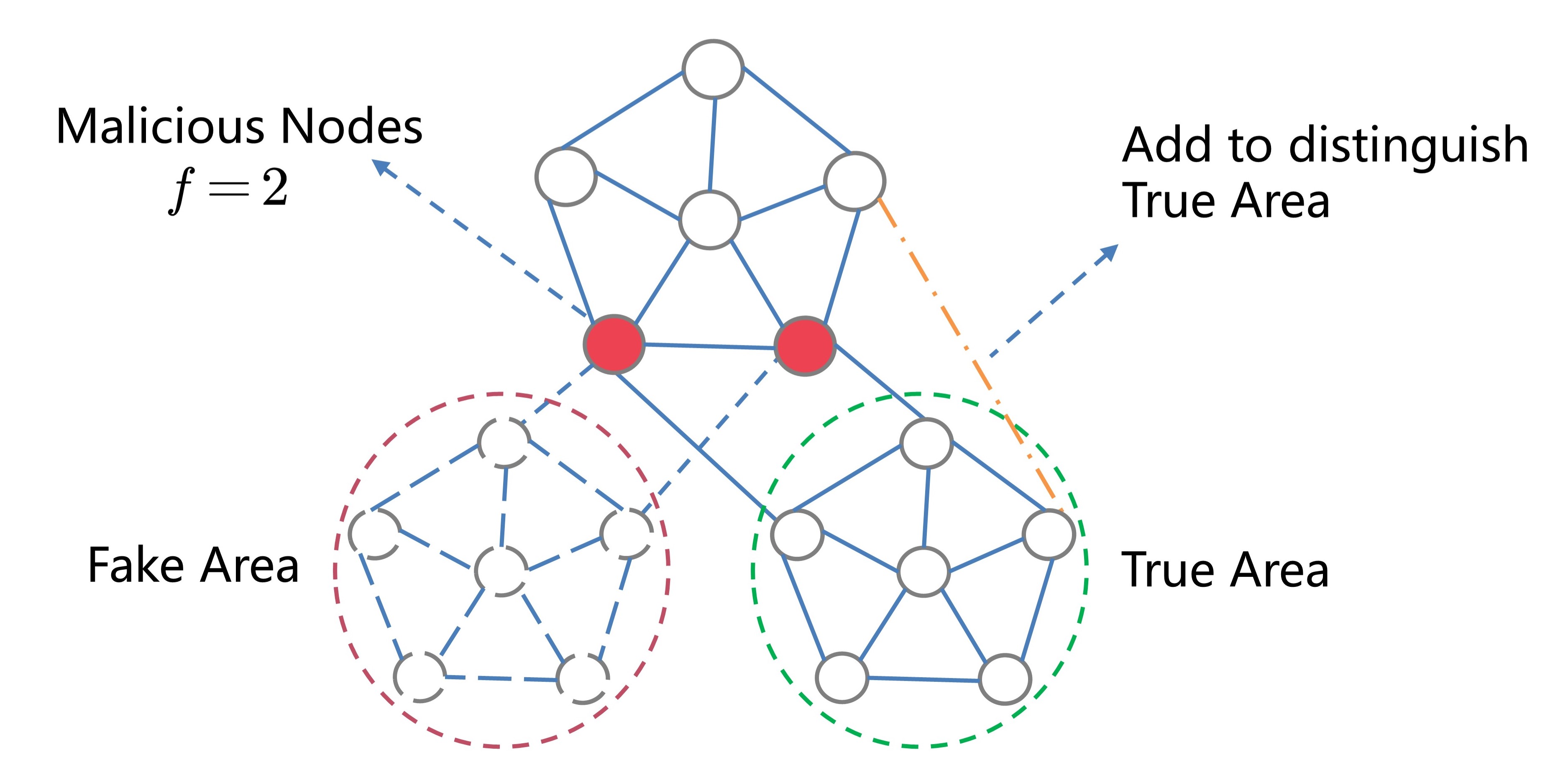}
\caption{An example of Eve faking an area in QKD networks. Red circle that represents Eve’s forged area, and a green circle that represents the real QKD network area, and both circles are similar in structure.}
\label{fig6}
\end{figure}

Next, to analyze a more general case of this type of attack, we consider that Eve can not only forge a fake node, but a fake area that is connected to malicious nodes in QKD network. As an example in figure \ref{fig6}, there are two malicious nodes (red nodes). The attacker wants to forge a fake area (red circle) to spoof other nodes in the network.

As shown in figure \ref{fig6}, there is a red circle that represents Eve’s forged area, and a green circle that represents the real QKD network area, and both circles are similar in structure. According to the verification process of our ITS distributed authentication scheme, when the orange connection does not exist, malicious nodes can replace the real region with a fake one. However, when the orange QKD link is added, the node connectivity from the green node to the real area is greater than the number of malicious nodes, indicating that there is always a path with honest nodes. Therefore, we require that the node connectivity should at least larger than the number of malicious nodes, and the length of key $k$ in any single QKD link used in the ITS distributed authentication should satisfy \eqref{eq17}.

\begin{equation}
\label{eq17}
k\ge \frac{-\log _2\varepsilon _{au}}{C-f}
\end{equation}

This setting because the length of $k$ determines the probability of the attackers guessing the right keys. In this setup, the probability of an attacker correctly guessing the key is not higher than $\varepsilon _{au}$.  For example, When node connectivity $C=f+1$, the success probability of the attacker is $ \varepsilon _{au} $.

\subsection{Non-repudiation}
This section we discuss the non-repudiation of our ITS distributed authentication. Non-repudiation requires that each node has a unique authentication key, and these keys are independent of each other.  Attackers can forge the same authentication key as a specific node to compromise non-repudiation. However, this issue has already been analyzed in section \ref{B1}. We are exploring another possibility for undermining non-repudiation. In this scenario, Eve selects a pair of nodes and utilizes controlled malicious nodes to make that these two nodes extract the same authentication key. Here, we are considering the attacker selecting a pair of adjacent nodes because some portions of the authentication keys for these two nodes are the same, making it easier for the attacker to achieve their goal. For any two neighboring nodes A and B, $ K_{raw}^{A} $  and $K_{raw}^{B} $ are partly identical since these two nodes have QKD links. We denote that $ K_{A\cap B}=K_{raw}^{A}\cap K_{raw}^{B} $ 

The independence of the two authentication keys of two nodes can also be expressed as mutual information $I_{AB} = I\left( K_{au}^{A}:K_{au}^{B},G_A,G_B \right)  $. According to lemma 1, \eqref{eq14} can be derived.
\begin{equation}
\label{eq14}
I_{AB} \le H\left( K_{au}^{A} \right) +\log _2\left( \frac{|K_{A\cap B}|}{|K_{raw}^{A}|}+\varepsilon _{PA} \right)   
\end{equation}

Given $ G_A,G_B\in \varepsilon _{PA}-ASU_2  $, if the original key lengths of nodes A and nodes B are equal,  we can derive that:
\begin{equation}
\label{eq15}
I\left( K_{au}^{A}:K_{au}^{B},G_A,G_B \right) =I\left( K_{au}^{B}:K_{au}^{A},G_A,G_B \right)    
\end{equation}
Here the length of $ K_{A\cap B} $ should be taken into consideration. Under the worst circumstance, the attackers can utilize the malicious nodes adjacent to A and B to increase $ |K_{A\cap B}| $. When there are $f/2$ malicious nodes distributed in the neighboring nodes of A and B respectively, we get $ |K_{A\cap B}|\le k+\frac{fk}{2} $. Similar to equation \eqref{eq12}, equation \eqref{eq14} can be written as:

\begin{equation}
\label{eq16}
I_{AB} \le \frac{2^{m+\lfloor f/2 \rfloor k+k-n}}{\ln 2} \le \frac{2^{-s}}{\ln 2} 
\end{equation}

We prove that $I_{AB}$ is less than the $I_{KE}$ in \eqref{eq12} (they are equal when $f=1$). If we choose the proper safety parameters $s$, the authentication keys of any two nodes are independent of each other even under the interference of the attackers.

\begin{figure*}[!t]
	\centering  
	\subfigure[ $\lambda=1.0$]{
		\label{fig7_first_case}
		\includegraphics[width=0.4\textwidth]{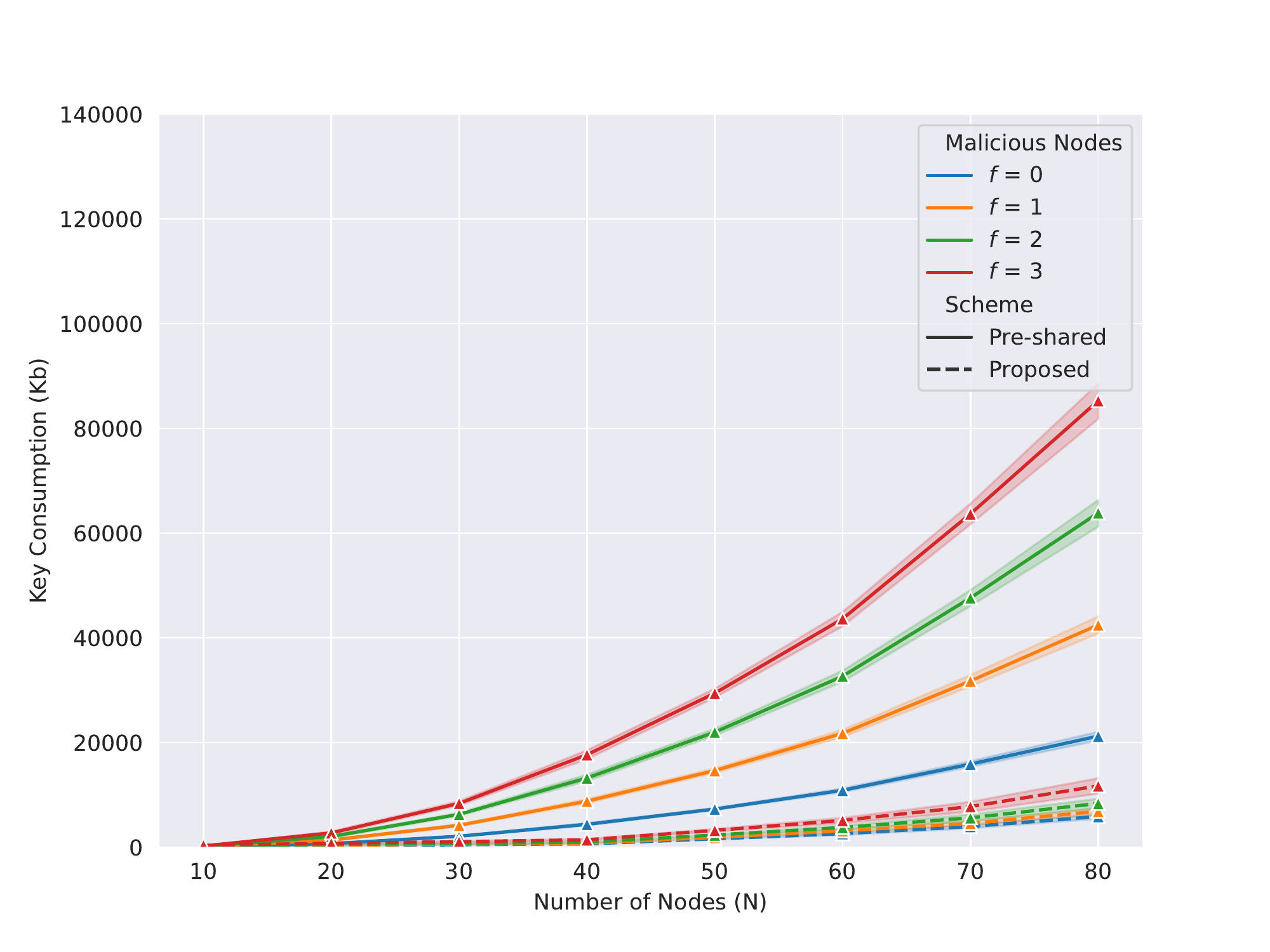}}
	\subfigure[ $\lambda=2.0$]{
		\label{fig7_second_case}
		\includegraphics[width=0.4\textwidth]{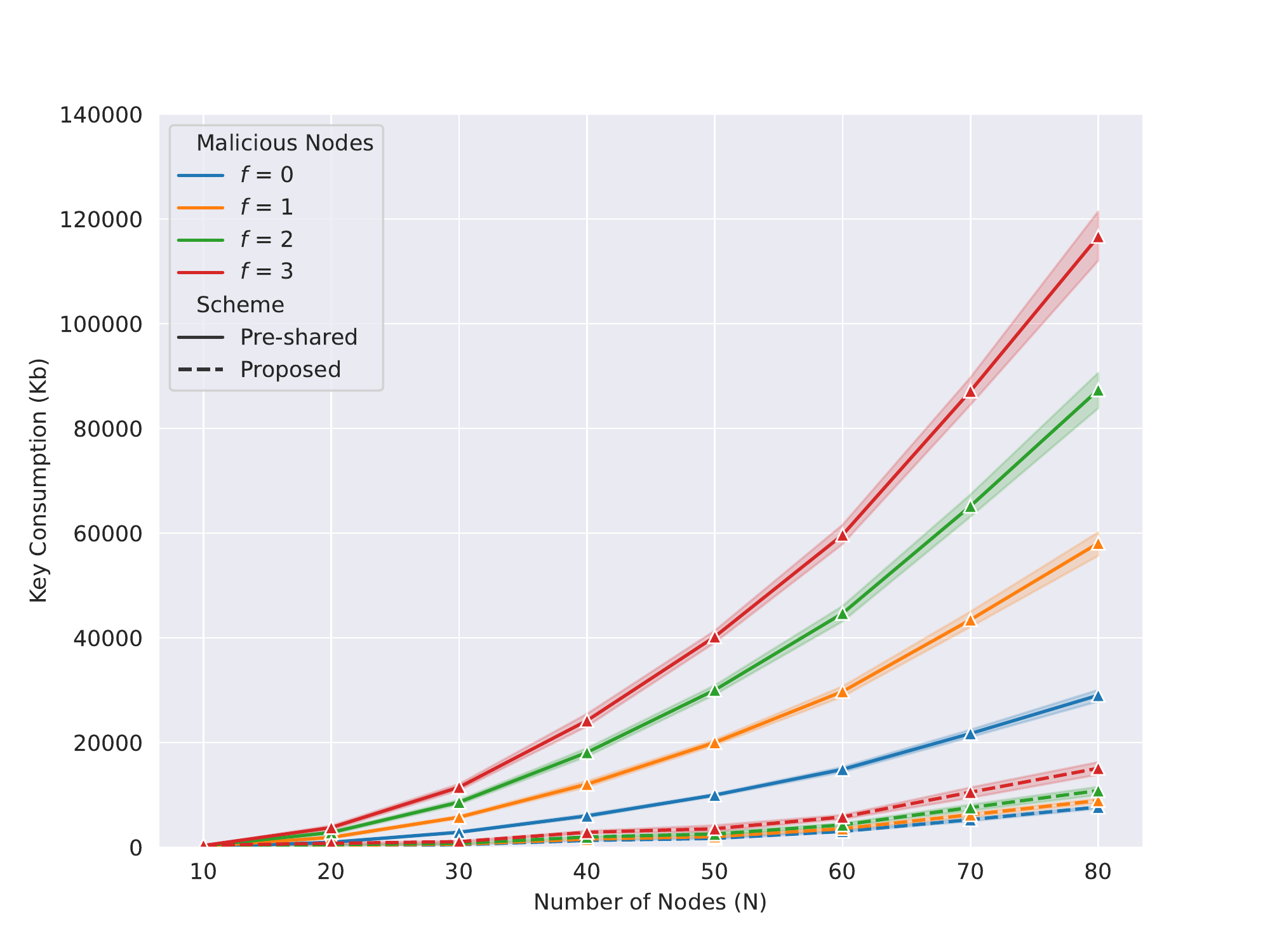}}
	\subfigure[ $\lambda=3.0$]{
		\label{fig7_third_case}
		\includegraphics[width=0.4\textwidth]{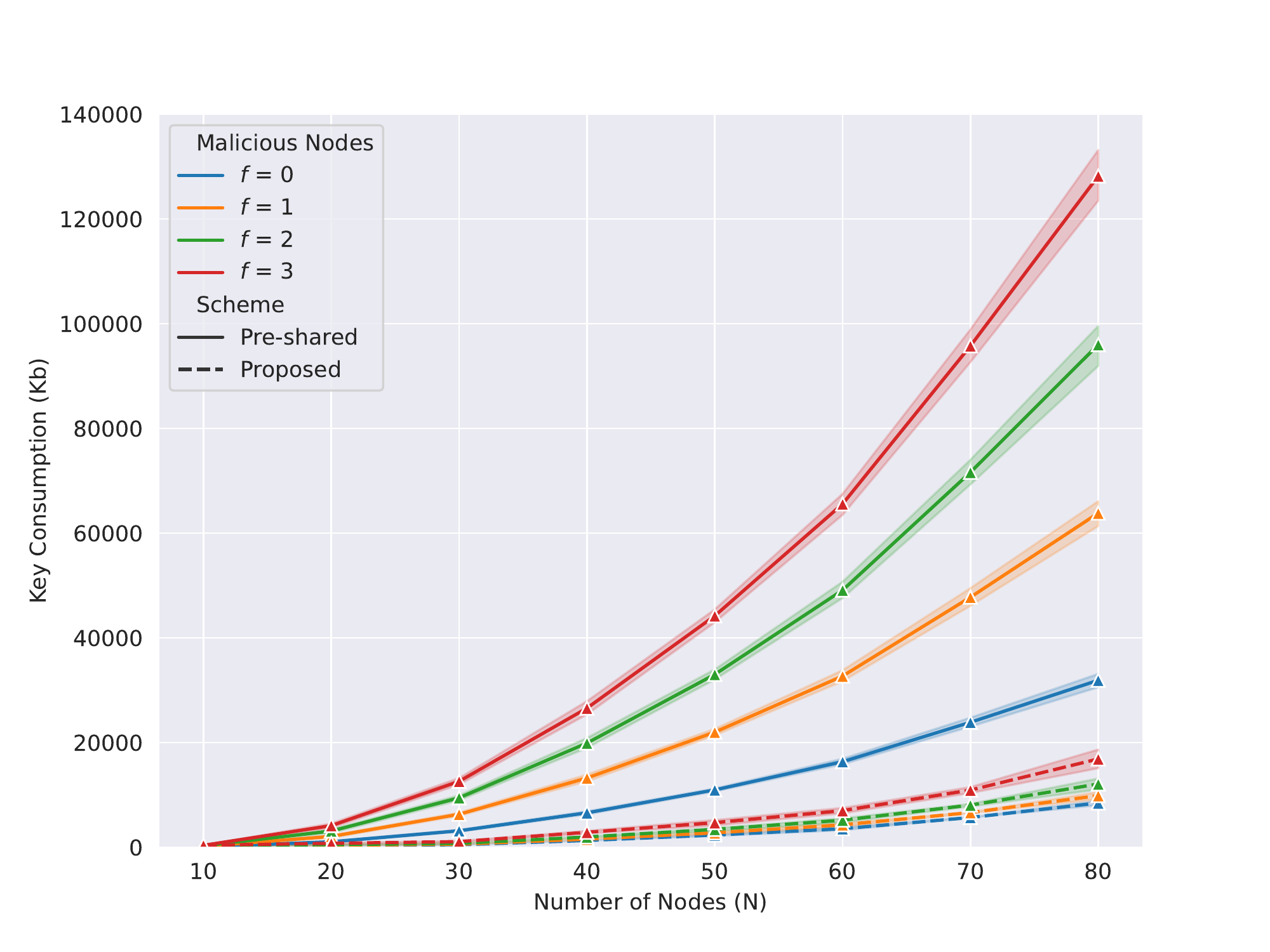}}
	\subfigure[ $\lambda=4.0$]{
		\label{fig7_fourth_case}
		\includegraphics[width=0.4\textwidth]{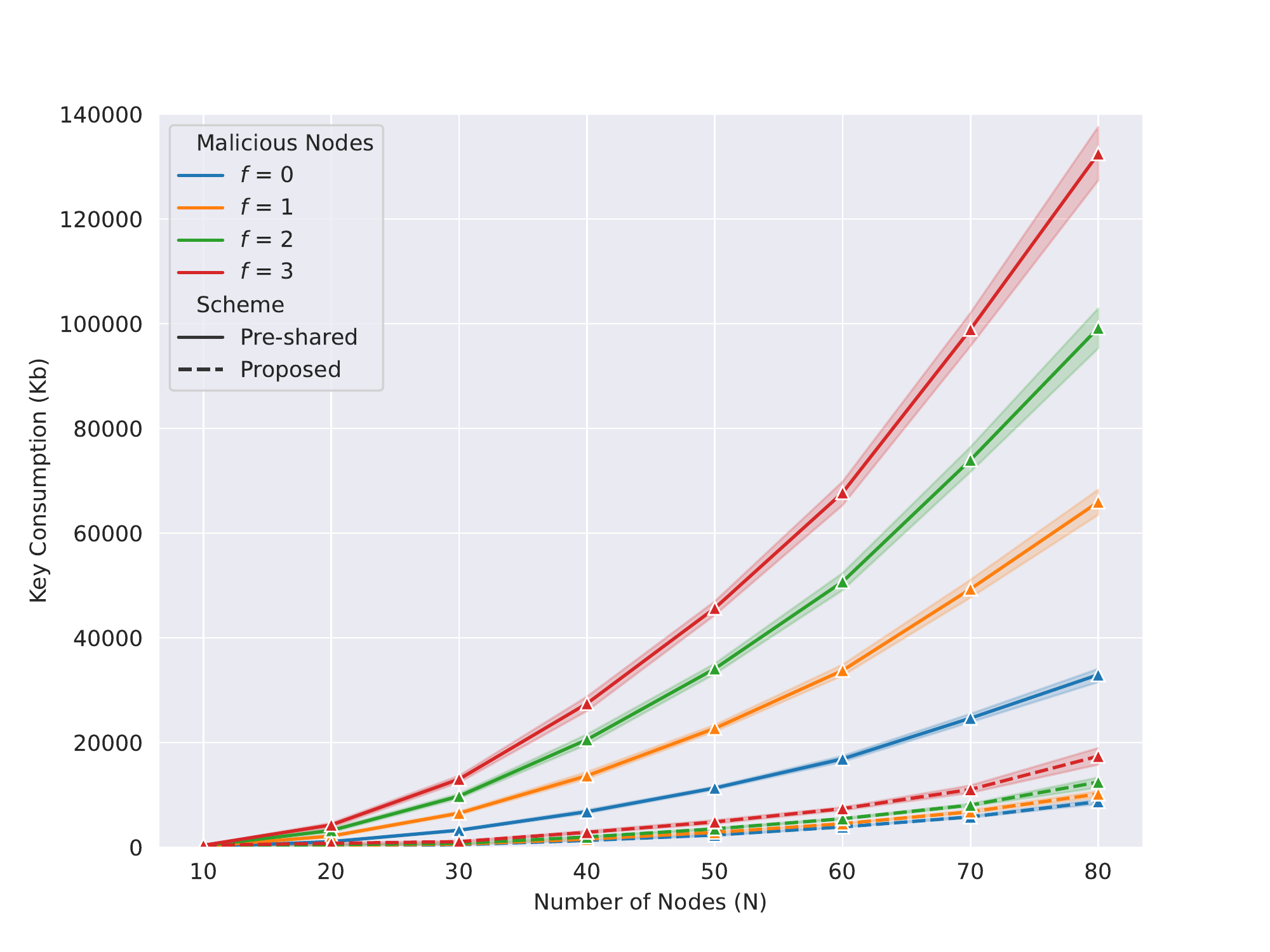}}
	\caption{Simulation between the consumption of point-to-point QKD keys (Kb) and the QKD network nodes $N$. 
		Figures \ref{fig7_first_case}- \ref{fig7_fourth_case} represent scenarios whrere the expectation of key distribution requirement $\lambda = {1.0, 2.0, 3.0, 4.0}$. Solid lines represent the pre-shared keys scheme, while dashed lines represent the proposed scheme.}
	\label{fig_sim_r2}
\end{figure*}

\section{Simulation}
We conducted simulations to assess the point-to-point key consumption of our scheme compared to the end-to-end pre-shared keys scheme under various scenarios involving different numbers of nodes, malicious nodes, and key distribution frequency. The end-to-end pre-shared keys can be equivalent to all the point-to-point QKD key consumption in the key distribution between these two nodes. This is because typically, the initial key is pre-shared, and subsequent keys are generated through key distribution.
\subsection{Simulation Setup}

QKD network setup: To obtain statistical simulation results under varying network topologies, we used a random network topology and conducted multiple experiments with a varying number of nodes ranging from 10 to 80. The malicious nodes nodes were set to {0, 1, 2, 3} in the simulations.  Simulation on different setup will be run 10 times with different random seeds, and the average values of these trials represent the simulation results \cite{Mehic2020Novel}. This can be used to more accurately verify the effectiveness of the  proposed scheme in this paper under different topologies.

Key distribution frequency setup :  The authentication scheme can aggregate all messages over a certain period of time, so the key consumption for authentication also depends on the frequency of sending message in key distribution process. To simulate the volatility of key distribution frequency, a poisson distribution was employed to model the frequency of packet transmission between any given pair of nodes \cite{Wang_Li_Han_Wang_2019}. Consequently, the times of key distribution frequency of a node pair were modeled as a poisson distribution $ d\left( \lambda \right)  $ with a mean value of $ \lambda /s $.  

Authentication setup: In both schemes, we utilized the method outlined in \cite{Kiktenko_Malyshev_Gavreev_Bozhedarov_Pozhar_Anufriev_Fedorov_2020} to construct universal hash function family, where the parameters are $ \omega =63,\varepsilon _{hash}=10^{-12} $. Based on the universal composability  \cite{Kiktenko_Malyshev_Gavreev_Bozhedarov_Pozhar_Anufriev_Fedorov_2020,Molotkov_2022},  $\varepsilon _{au} =\varepsilon _{hash} + \varepsilon _{PA}$. Additionally, in the pre-shared keys scheme, we applied key recycling mechanism \cite{Portmann_2014}. The PA scheme we used in ITS distributed authentication is Toeplitz Hash \cite{Krawczyk_1994} with $\varepsilon _{PA}=10^{-12}$ . To ensure fairness in the comparison, in our proposed scheme, we account for all authentication key consumption during the consensus process. Additionally, to mitigate the impact of aggregating information, in the pre-shared keys scheme, we have also incorporated a mechanism for aggregating information. The pre-shared keys scheme will aggregate and authenticate messages within every $\varDelta$ time period. Here, $\varDelta$ is equal to 1s (seconds).

\begin{figure*}[!t]
\centering  
\subfigure[Malicious nodes number $f=0$]{
\label{fig6_first_case}
\includegraphics[width=0.4\textwidth]{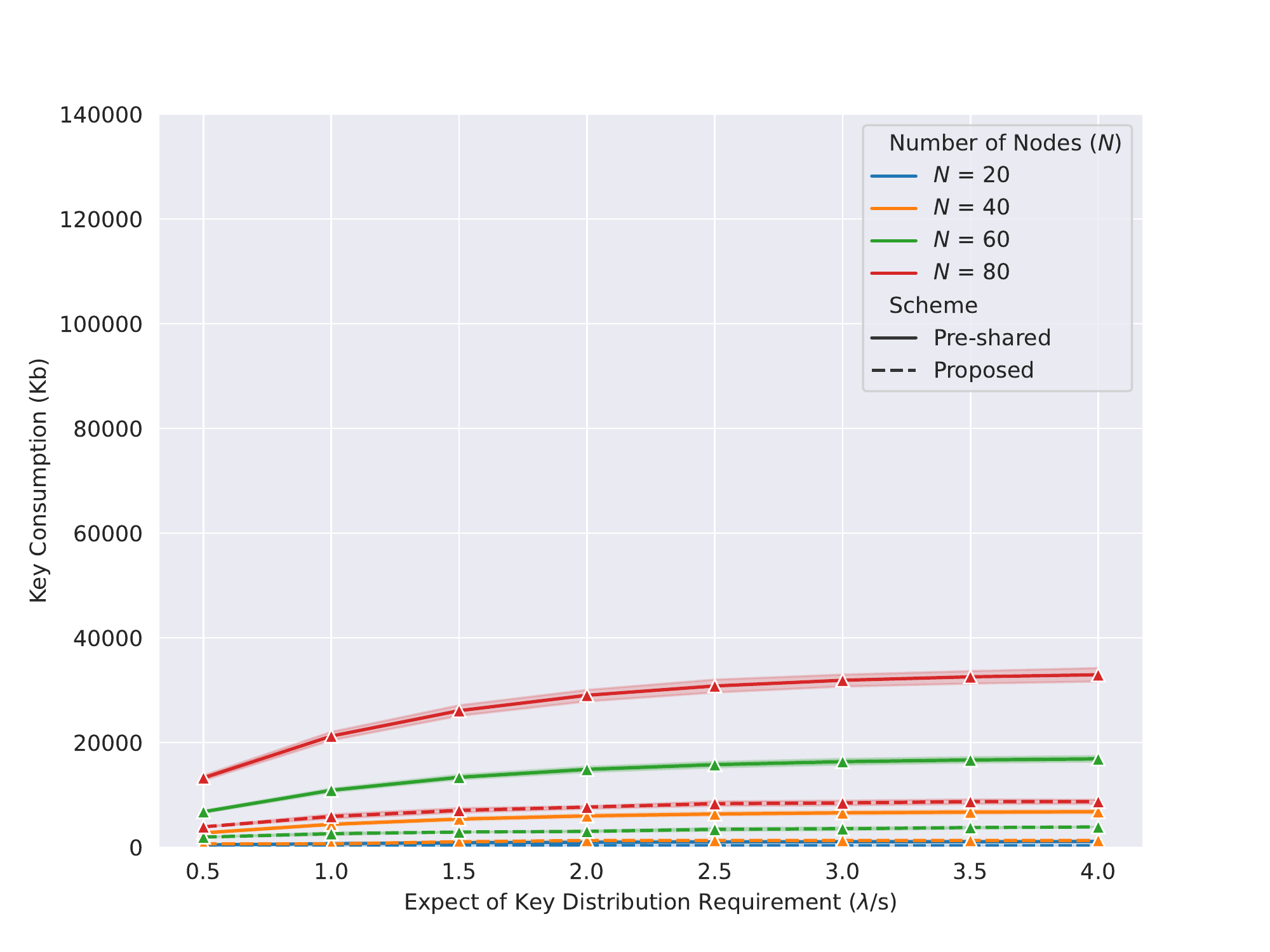}}
\subfigure[Malicious nodes number $f=1$]{
\label{fig6_second_case}
\includegraphics[width=0.4\textwidth]{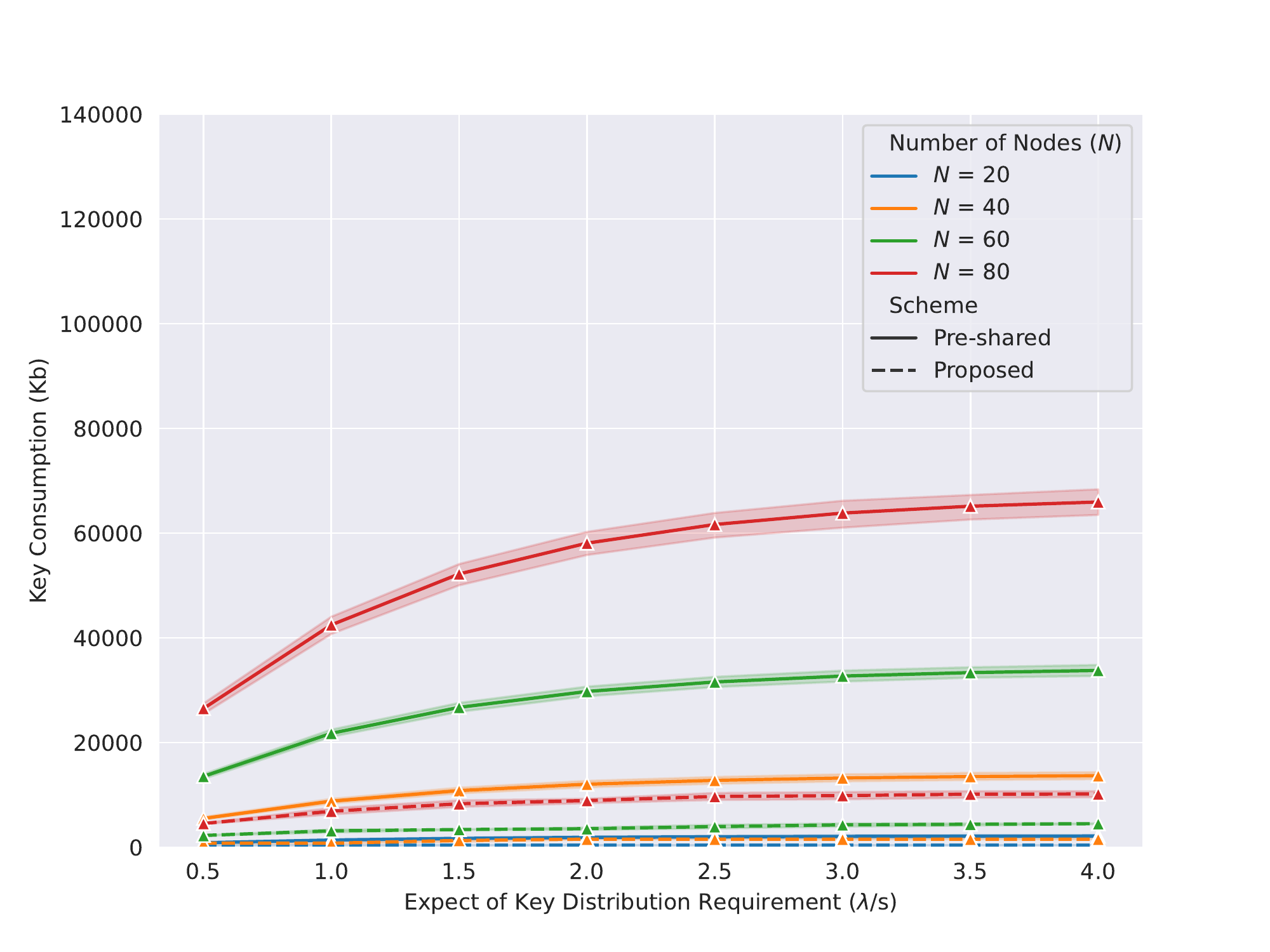}}
\subfigure[Malicious nodes number $f=2$]{
\label{fig6_third_case}
\includegraphics[width=0.4\textwidth]{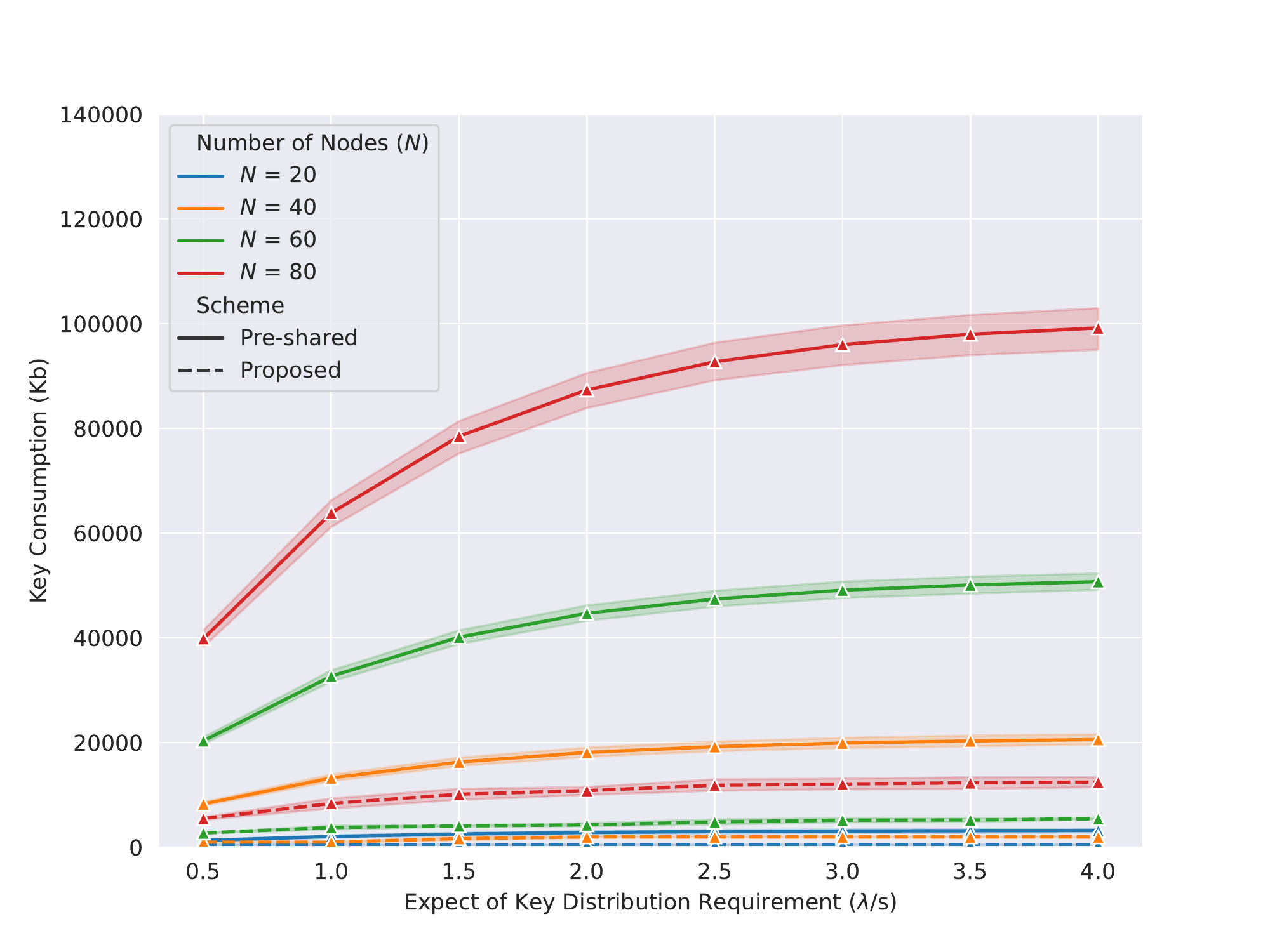}}
\subfigure[Malicious nodes number $f=3$]{
\label{fig6_fouth_case}
\includegraphics[width=0.4\textwidth]{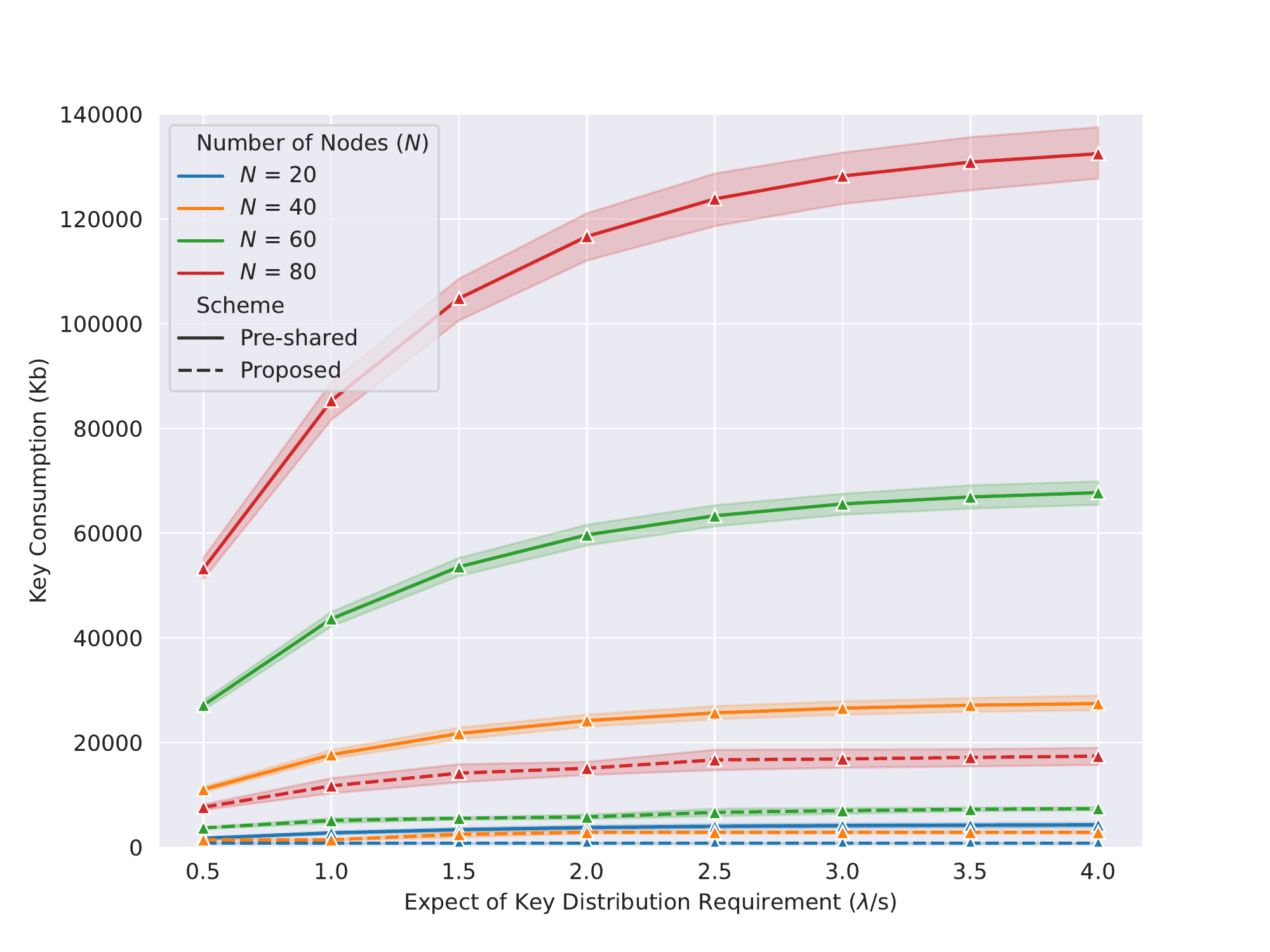}}
\caption{Simulation between the consumption of point-to-point QKD keys (Kb) and the expected value of key distribution frequency $\lambda$. 
Figures \ref{fig6_first_case}-\ref{fig6_fouth_case} represent scenarios with different malicious nodes $f = {1, 2, 3, 4}$. Solid lines represent the pre-shared keys scheme, while dashed lines represent the proposed scheme.}
\label{fig_sim_r1}
\end{figure*}

\subsection{Simulation Results}
Our simulation results are primarily divided into two parts: one focuses on key consumption with respect to the number of nodes N, while the other examines key consumption in relation to key distribution frequency $\lambda$.

As shown in Figure \ref{fig_sim_r2}, We demonstrated the relationship between the point-to-point key consumption and the number of nodes for both schemes at different values of $\lambda = {1.0, 2.0, 3.0, 4.0}$. In the pre-shared keys scheme, the rate of key consumption growth with an increasing number of network nodes $N$ far exceeds that of our proposed scheme. For example, Figure \ref{fig7_first_case}, when $\lambda = 1.0$ and $f = 1$, our scheme's key consumption is only 13.1\% of the pre-shared keys scheme. In the pre-shared keys scheme, the rate of key consumption growth with an increasing number of network nodes indeed surpasses that of our proposed scheme. For instance,  in figure \ref{fig7_first_case}, when $\lambda = 1.0$ ,$f = 1$ and $N =10$, our scheme's key consumption is 151.3\% of the pre-shared keys scheme. However, when $N$ increase to 80, our scheme's key consumption is only 16.2\% of the pre-shared keys scheme. Different values of $f$ will also affect the amount of key consumption, with higher values of $f$ leading to greater key consumption. In figure \ref{fig7_fourth_case}, when $N = 80$, $\lambda = 4.0$  and $f$ takes on values of 0, 1, 2, and 3, our scheme's key consumption is 26.5\%, 15.4\% , 13.5\% and 13.1\% of the pre-shared keys scheme, respectively.

The key consumption for both schemes is also dependent on the value of $\lambda$. Due to the ability of the authentication scheme to aggregate messages within the same time frame, the key consumption rate increases more slowly with the growth of $\lambda$, which is slower compared to the quadratic rate at which it increases with the number of nodes $N$. For example, When $N = 80$, $f=1$ and $\lambda$ takes on values of 1.0, 2.0, 3.0 and 4.0, our scheme's key consumption is 16.2\%, 15.37\% , 15.46\% and 15.43\% of the pre-shared keys scheme, respectively.

%
%Table \ref{table1} displays the simulation results for different topologies and different numbers of malicious nodes. Our simulation results explore the relationship between the key consumption and the properties of the network for both schemes in the presence of malicious nodes. 
\subsection{Complexity Analysis of Key Consumption}
In the previous end-to-end pre-shared keys scheme, due to the a key recycling mechanism, the length of the end-to-end key required for each authentication is the same as the length of the tag. This portion of the key needs to be generated with the assistance of point-to-point QKD key. Its key consumption is related to the number of nodes and the path length. First we consider the key consumption between two non-adjacent nodes $p$ and $q$ as \eqref{eq19}. 

\begin{equation}
\label{eq19}
C_{pq}=|tag|\sum_{z=1}^{f+1}{P_z\left( p,q \right)} 
\end{equation}

Here we consider the case of distribution along $f+1$ disjoint paths. In this case malicious nodes eavesdrop on the least number of keys, resulting in less key consumption. ${P_z\left( i,j \right)}$ represents the lengths of the $z$th paths between the node $p$ and $q$.

In general, the key needs to be pre-shared between any two non-adjacent $p$ and $q$, noted as $p\notin adj\left( q \right)$. As a result, the key consumption of the pre-shared keys scheme in total network $C_{pre}$ is:

\begin{equation}
\label{eq20}
C_{pre}=\sum_{p\in V}{\sum_{q\in V}{C_{pq}}},   p\notin adj\left( q \right)    
\end{equation}

Here let $d$ be the average path length of the network, we can derive that the consumption of pre-shared keys is about $ O\left( (f+1)N^2d \right)  $ level.

However, in our proposed scheme, the key consumption is closely related to the edges of the network. If we let $k_{max}$ represent the maximum value of the number of keys extracted from the QKD link, then the maximum value of the total network key consumption in consensus $C_{con}$ is:

\begin{equation}
\label{eq18}
C_{con} = Ek_{\max}\cdot step
\end{equation}

where $E$ is the number of network edges and $step$ is the number of broadcast in all the consensus process. Since $k_{max}$ is smaller than the constant number $|K_{au}|$, the complexity of key consumption in the consensus scheme is $ O\left( E \cdot step \right) $.
%Our simulation results verify the above key consumption complexity. When the network size is small, as figure \ref{fig_first_case}, the key consumption of the consensus scheme will be 4 times larger than that of the pre-shared key scheme. However, the key consumption growth rate of consensus scheme with the network size is much smaller than that of the pre-shared key scheme. For example, in figure \ref{fig_second_case}, the key consumption of the consensus scheme is already smaller than that with the pre-shared key scheme. In figure \ref{fig_third_case}, the key consumption of the consensus scheme is only 16.71\%  of the pre-shared key scheme. Considering the future growth of network size, the consensus scheme is more advantageous than the pre-shared key scheme.

%\nocite{*}

\section{Conclusion an Discussion}
In this paper, we propose an new QKD network scheme that enables end-to-end key distribution against $ MIN\left( C-1,\lfloor \frac{N-1}{2} \rfloor \right)  $ malicious nodes. Our proposed solution fully takes into account the differences in security requirements between point-to-point QKD systems and QKD networks. Our scheme supplements two important fundamental properties in the authentication issue: identity unforgeability and non-repudiation. It exhibits a significantly lower growth trend in key consumption and does not require pre-shared end to end keys.
%In addition, we propose a BFT signature scheme that does not require pre-shared keys to ensure the information-theoretic security of the consensus process.

From the perspective of distributed systems, multiple nodes are required to establish trust relationships and cooperate with each other in QKD networks. Considering that there are malicious nodes among cooperators, the security issues we need to consider are far more complicated than a stand-alone QKD system. Our proposed framework provides a general way to handle problems that require multiple nodes to cooperate in QKD networks. For other applications in QKD networks, we only need to modify the proposal content in the consensus process. These modifications will not affect the security and correctness of this framework. Therefore, this scheme could potentially be further applied to handle routing \cite{WOS:000943552900001} or complex resource allocation \cite{WOS:001062432100003} tasks in QKD networks.

In another perspective, our consensus scheme and ITS distribution authentication scheme can reflect whether the current state of the QKD networks satisfies the security conditions. We find that the security of a QKD network is related to its connectivity and the number of honest nodes. We can determine whether the network has the ability to satisfy the security conditions from the number of valid authentication results and the connectivity of the key graph in consensus. This can be used as a preventive mechanism, just like quantum bit error rate (QBER) in QKD systems. We will explore it further in the future.

%First, we proposes ITS fault-tolerance-consensus process to resist active attacks and passive attacks in the presence of malicious nodes. Second, we propose the BFT signature to guarantee the information-theoretical security of the consensus scheme. We analyze the advantages of our network architecture, which guarantees not only the ITS under active and passive attacks but also the fairness and BFT ability. Our simulations demonstrate key consumption of the consensus scheme and its growth trend with QKD networks

%As a conclusion, ITSBFT-QKD network scheme can effectively improve the fault tolerance ability of QKD networks and provide the highest level of security in currently relay-based QKD networks. Therefore, the proposed scheme could be applied to other applications(e.g.  Resource allocation in QKD networks, Quantum blockchain, Quantum Key agreement and Secret sharing) in the future and has the potential to become the next-generation security infrastructure.
Finally, the scheme we propose has the potential to be integrated with MDI or TF QKD networks. Such integration not only can minimizes the need for trusted relays but also accommodate untrusted nodes. Given the inherent characteristics of MDI or TF QKD, we are optimistic that this approach will be resistant to the risks posed by malicious nodes in the future. Dispelling doubts about the security of relay-based QKD networks, it will significantly promote the application of QKD networks.

\section*{Acknowledgment}
We sincerely thank Professor Hoi-Kwong Lo for his invaluable insights and constructive discussions. 
We acknowledge that this work is supported by the National Natural Science Foundation of China (grant number: 62071151).

\ifCLASSOPTIONcaptionsoff
\newpage
\fi



%\bibliographystyle{IEEEtranTCOM}

\bibliographystyle{IEEEtran}
\bibliography{bibfile}

\end{document}